\documentclass[12pt,letterpaper]{article}

\usepackage{amsmath}
\usepackage{amssymb}
\usepackage{epsfig}
\usepackage{graphicx}
\usepackage{cite}

\newcommand{\capdef}{}
\newcommand{\mycaption}[2][\capdef]{\renewcommand{\capdef}{#2}%
       \caption[#1]{{\footnotesize #2}}}
\makeatletter
\renewcommand{\fnum@table}{\textbf{\tablename~\thetable}}
\renewcommand{\fnum@figure}{\textbf{\figurename~\thefigure}}
\makeatother

\newcounter{myenumi}

\renewcommand{\themyenumi}{\roman{myenumi}}
{\end{list}}

\newlength{\myem}
\newcounter{mysubequation}[equation]

\makeatletter
\renewcommand{\section}{\@startsection{section}{1}{0em}{-\baselineskip}%
{\baselineskip}{\normalfont\large\bfseries}}
\renewcommand{\subsection}%
{\@startsection{subsection}{2}{0em}{-0.7\baselineskip}%
{0.7\baselineskip}{\normalfont\bfseries}}
\makeatother

\newcommand{\ie}{{\it i.e.}}

\newcommand{\eg}{{\it e.g.}}

\newcommand{\cf}{{\it cf.}}

\newcommand{\eq}{Eq.}

\newcommand{\fig}{Figure}

\newcommand{\Ref}{Ref.}
\newcommand{\Refs}{Refs.}

\newcommand{\App}{Appendix}

\newcommand{\Tab}{Table}

\newcommand{\JHFHK}{\mbox{{\sf T2HK}}$^*$}
\newcommand{\NuFactII}{\mbox{\sf NuFact-II}}

\newcommand{\stheta}{\sin^22\theta_{13}}
\newcommand{\deltacp}{\delta_\mathrm{CP}}
\newcommand{\ldm}{\Delta m_{31}^2}
\newcommand{\sdm}{\Delta m_{21}^2}
\newcommand{\equ}[1]{\eq~(\ref{equ:#1})}
\newcommand{\figu}[1]{\fig~\ref{fig:#1}}
\newcommand{\tabl}[1]{\Tab~\ref{tab:#1}}
\newcommand{\bi}{\begin{itemize}}
\newcommand{\ei}{\end{itemize}}

\begin{document}

\begin{titlepage}

\renewcommand{\thefootnote}{\alph{footnote}}

\vspace*{-3.cm}
\begin{flushright}
TUM-HEP-593/05\\
MADPH-05-1428
\end{flushright}

\vspace*{0.5cm}

\renewcommand{\thefootnote}{\fnsymbol{footnote}}
\setcounter{footnote}{-1}

{\begin{center}
{\Large\bf Physics and optimization of 
beta-beams:\\ 
From low to very high
gamma}
\end{center}}
\renewcommand{\thefootnote}{\alph{footnote}}

\vspace*{.8cm}
{\begin{center} {\large{\sc
                P.~Huber\footnote[1]{\makebox[1.cm]{Email:}
                phuber@physics.wisc.edu},~
                M.~Lindner\footnote[2]{\makebox[1.cm]{Email:}
                lindner@ph.tum.de},~
                M.~Rolinec\footnote[2]{\makebox[1.cm]{Email:}
                rolinec@ph.tum.de},~
                W.~Winter\footnote[3]{\makebox[1.cm]{Email:}
                winter@ias.edu}
                }}
\end{center}}
\vspace*{0cm}
{\it
\begin{center}

\footnotemark[1]
       Department of Physics, University of Wisconsin, \\
       1150 University Avenue, Madison, WI 53706, USA

\vspace*{1mm}

\footnotemark[2]%
       Physik--Department, Technische Universit\"at M\"unchen, \\
       James--Franck--Strasse, 85748 Garching, Germany

\vspace*{1mm}

\footnotemark[3]%
       School of Natural Sciences, Institute for Advanced Study, \\
       Einstein Drive, Princeton, NJ 08540, USA

\vspace*{1cm}

\today
\end{center}}

\vspace*{0.3cm}

\begin{abstract}
The physics potential of beta beams is investigated from low 
to very high gamma values and it is compared to superbeams 
and neutrino factories. The gamma factor and the baseline 
are treated as continuous variables in the optimization of 
the beta beam, while a fixed mass water Cherenkov detector 
or a totally active scintillator detector is assumed.
We include in our discussion also the 
gamma dependence of the number of ion decays per year.
For low gamma, we find that a beta beam could be a very 
interesting alternative to a superbeam upgrade, especially 
if it is operated at the second oscillation maximum to 
reduce correlations and degeneracies. For high gamma, we 
find that a beta beam could have a potential similar to 
a neutrino factory. In all cases, the sensitivity of the 
beta beams to CP violation is very impressive if similar
neutrino and anti-neutrino event rates can be achieved.
\end{abstract}

\vspace*{.5cm}

\end{titlepage}

\newpage

\renewcommand{\thefootnote}{\arabic{footnote}}
\setcounter{footnote}{0}


\section{Introduction}

Future neutrino oscillation experiments with beams will
ultimately lead to precise measurements of the leading 
solar and atmospheric oscillation parameters. This will 
also lead to precise measurements of $\stheta$, to a 
determination of the mass hierarchy, and measurements 
of leptonic CP violation. Long baseline neutrino 
oscillation experiments constitute therefore a very 
valuable physics program, since they will provide the 
most precise information on flavour which will be crucial
for progress in our understanding of the origin and the
structure of flavour. 
There exist different experimental paths how precision 
oscillation measurements can be achieved. Beyond the 
planned superbeam neutrino oscillation experiments 
T2K  \cite{Itow:2001ee}  and NO$\nu$A \cite{Ayres:2004js}), 
one could build upgrades of superbeams 
(T2HK \cite{Nakamura:2003hk}), neutrino factories 
\cite{Albright:2000xi,Blondel:2000gj,Apollonio:2002en}, 
or $\beta$-beams
\cite{Zucchelli:2002sa,Bouchez:2003fy,Mezzetto:2003ub,Burguet-Castell:2003vv,Terranova:2004hu,Donini:2004hu,Mezzetto:2004gs,Albright:2004iw,Donini:2004iv,Burguet-Castell:2005pa,Agarwalla:2005we,Donini:2005rn}.
The optimization among these alternatives depends on 
the outcome of on-going experiments which determine 
or limit, for example, the actual value of $\stheta$, 
which determines the CP violation and mass hierarchy 
measurement reaches. More important are, however, 
the generic differences between the different options. 
Since superbeam upgrades are limited by the intrinsic
beam background, their performance becomes rapidly
suppressed below $\stheta \lesssim 10^{-3}$. Contrary 
to that, neutrino factories have a clean beam which 
contains 50\% neutrinos and 50\% anti-neutrinos of 
different flavours, which means that they are not 
limited by an intrinsic beam background. However,
neutrino factory detectors must be able to discriminate 
extremely well between right-sign and wrong-sign muons
to discriminate neutrinos and anti-neutrinos~\cite{Cervera:2000kp}.
This can be done with magnetized detectors, but it 
heavily affects the efficiencies especially at low energies,
limiting the accessible L/E range. In addition, for 
a $\stheta$ or CP violation search experiment, high event 
rates turn out to be very important, which means that 
combinations of high neutrino energies and quite long 
baselines are favored (see, e.g. \cite{Freund:2000ti}).
$\beta$-beams, on the other side, face neither of these 
principle disadvantages: They use a pure electron neutrino 
(anti-neutrino) beam and they can be easily operated at 
the oscillation maximum at a rather reasonable baseline. 
Besides there also is a proposal for a very low $\gamma\sim10$ 
$\beta$-beam which could be used to study neutrino nucleon 
interactions~\cite{Volpe:2003fi,Serreau:2004kx}. Although the 
insights gained at such a facility would be invaluable to improve 
our theoretical understanding of neutrino cross sections, we will not 
discuss this option any further as it would be beyond the scope of this work.

All these novel types of experiments have certainly a number
of technological issues which must be further investigated 
before such facilities can be built. However, the physics 
questions, such as the optimal $\gamma$ and baseline combinations
for $\beta$-beams, should be understood first in order to 
identify the best options which we should be aiming for.

The answers to such questions depends obviously on the chosen detector
technology, external constraints to the setup, and the physical
observable for which the optimization of the $\beta$-beam is
performed.  Some of these issues have been previously discussed in
detail with a somewhat different focus
in~\cite{Burguet-Castell:2003vv,Burguet-Castell:2005pa}.

This study addresses the most relevant of these questions.
We analyze in detail the optimization of $\beta$-beams
and we compare the physics potential of $\beta$-beams to 
the one of superbeam upgrades and neutrino factories, where
a special emphasis is put on an ``equal footing'' comparison, 
\ie, we choose comparable detector sizes and running times.

This study is organized as follows: 
In chapter \ref{sec:detector}, we describe the simulation 
techniques and the parameters used for the beam and for the 
detectors of this study. The optimization of the $\stheta$ 
sensitivity is then discussed in chapter~\ref{sec:optimals13}.
Special emphasis is here put on the dependence on $\gamma$,
the baseline and the $\gamma$ scaling of the number of decays per year.
Chapter~\ref{sec:optimalHCP} contains then a discussion
of the optimization with respect to the mass hierarchy and 
CP violation. A summary and conclusions will be given finally
in section~\ref{sec:summary}.

\section{Experiment simulation}
\label{sec:detector}

In this section, we first describe the properties of the 
$\beta$-beam and the techniques used for the simulation 
of the experiment. One of the main objectives of this
study is to investigate the performance of the $\beta$-beam
in a wide range of $\gamma$. Different choices of $\gamma$ 
imply different optimal detector technologies and the 
description of the detector parameterizations is therefore 
a key element of this section.

\subsection{Simulation techniques}

All experiment simulations in this study are performed 
with the GLoBES software \cite{Huber:2004ka}. The assumed 
``true'' solar and atmospheric oscillation parameters, 
which are used as input for the calculation of simulated 
event rates with GLoBES, are, unless stated otherwise
\cite{Fogli:2003th,Bahcall:2004ut,Bandyopadhyay:2004da,Maltoni:2004ei}:
\begin{equation}
\begin{array}{ccc}
\ldm = 2.5 \, \cdot 10^{-3} \, \mathrm{eV}^2, & & \sin^2 2 \theta_{23}=1 \\
 & & \\
\sdm = 8.2 \cdot 10^{-5} \, \mathrm{eV}^2, & & \sin^2 2 \theta_{12}=0.83.
\end{array}
\label{equ:standard_params}
\end{equation}
The errors of these parameters are given in
\Refs~\cite{Fogli:2003th,Bahcall:2004ut,Bandyopadhyay:2004da,Maltoni:2004ei}
will also be included in our analysis. For $\stheta$, we only
allow values below the CHOOZ bound~\cite{Apollonio:1999ae}, \ie,
$\stheta \lesssim 0.1$. In some cases, we additionally demonstrate
the effects for the choice of a smaller true value of
$\ldm = 1.5 \cdot 10^{-3} \, \mathrm{eV}^2$, representing the lower
end of the currently allowed 90\% CL region~\cite{Maltoni:2004ei}
and, how this affects the performance of the simulated experiments.

The solar oscillation parameters and their errors are included as
external input which affect the performance of
the discussed setups via correlations. We assume a precision of 5\% for $\sdm$
and 10\% for $\theta_{12}$. This corresponds approximately to the
current precisions such that this should be a conservative assumption
at the time of the discussed experiments~\cite{Bahcall:2004ut}.
In addition, $\beta$-beams cannot determine the leading atmospheric
parameters with high precision. Therefore, we combine all of our
$\beta$-beam simulations, which include effects due to correlations
with the atmospheric parameters, with an equivalent of 10 years of
T2K running\footnote{See \Refs~\cite{Huber:2002mx,Huber:2002rs} for details of
the T2K description within GLoBES.}. Data corresponding to this
scenario in their combined statistics should be available from
T2K, NO$\nu$A, and atmospheric experiments at the time when a 
$\beta$-beam is analyzed. However, we only use the T2K 
disappearance channels (and leave out the appearance channels) 
in order to avoid confusion in the interpretation of $\beta$-beam 
appearance data. This assures that we only include external 
information on the leading atmospheric parameters. 
This approach may seem to be more complicated than assuming 
external precisions for the leading atmospheric parameters. 
However, if one discusses the effects of parameter degeneracies, 
it is always a difficult issue {\em where} (at which point in 
parameter space) these external precisions for the degenerate 
solutions should be centered, \ie\ where the external measurement
gives the degenerate solution. Thus, simply assuming external
precisions added at some points in parameter space would clearly 
over-estimate the performance of the $\beta$-beams by adding
unwanted prior contributions. Including in the analysis an
equivalent of 10 years of T2K disappearance running is therefore
a simple and quite realistic approach to deal with this problem 
without creating ``artificial'' topologies. For the neutrino 
factory and T2HK simulations, however, it is reasonable to assume
that $\ldm$ and $\theta_{23}$ are measured by far best with 
their own disappearance channels, \ie, we do not impose any 
external precision on the leading atmospheric parameters there. 
Eventually, we assume a constant matter density profile with 
a 5~\% uncertainty  on the value of the baseline-averaged 
matter density, where the uncertainty takes into account matter 
density uncertainties as well as matter density profile 
effects~\cite{Geller:2001ix,Ohlsson:2003ip,Pana}. If the 
baseline of an experiment is changed, we also re-compute and 
use the average matter density for this baseline. 

\subsection{Beam characteristics}

The neutrino beam discussed in this study is assumed to originate
from the beta decay of $^6$He and $^{18}$Ne isotopes in straight
sections of a storage ring. It is assumed that the energy or
equivalently the relativistic $\gamma$ value can be chosen.
The corresponding decay channels are:
\begin{equation}
\begin{array}{ccc}
^{18}Ne & \rightarrow & ^{18}F + e^+ + \nu_e \\ \\
^{6}He & \rightarrow  & ^{6}Li + e^- + \bar{\nu}_e \\
\end{array}
\end{equation}
which leads correspondingly to pure electron neutrino or
electron anti-neutrino beams without any intrinsic beam
contamination. In the rest frame of the particular decay
the neutrinos are emitted isotropically and the energy
spectrum is given by the well-known beta decay with a
certain endpoint energy for a given isotope. In the storage
ring, the spectrum is boosted and it becomes in the
laboratory system~\cite{Burguet-Castell:2003vv}:
\begin{equation}
\frac{d\phi}{dE_\nu} \propto
\frac{E_\nu^2}{\gamma}
\left( 1-\frac{E_\nu}{2 \gamma (E_0 + m_e)} \right)
\sqrt{
 \left( 1-\frac{E_\nu}{2 \gamma (E_0 + m_e)} \right)^2 -
  \left( \frac{m_e}{E_0 + m_e} \right)^2
}.
\end{equation}

$^{18}$Ne and $^6$He are in principle not the only possible
choices for the isotopes, but we will use here the same
assumption as most of the existing literature~\cite{Bouchez:2003fy,Mezzetto:2003ub,Burguet-Castell:2003vv,Terranova:2004hu,Donini:2004hu,Mezzetto:2004gs,Donini:2004iv,Burguet-Castell:2005pa,Agarwalla:2005we,Donini:2005rn}.
The endpoint energies $E_0$ of these two isotopes are very similar.
This leads to the nice feature, that they give approximately
the same mean neutrino energy at the same $\gamma$, \ie we
have $\gamma(^6He) = \gamma(^{18}Ne)$ and there is no obvious
gain in increasing one of the two $\gamma$'s
(\cf, \Ref~\cite{Burguet-Castell:2005pa})\footnote{Basically it is
fortunate to have a $\gamma$ as high as possible
for each neutrinos and anti-neutrinos independently.}.
The endpoint energies are $E_0=3.4\,\mathrm{MeV}$ for
$^{18}$Ne and $E_0=3.5\,\mathrm{MeV}$ for $^6$He.
Note that we neglect the fact that there are different
exited states in the daughter nuclei of the decay, which
additionally lead to negligible small contributions to the 
spectra with different endpoint energies.

In \Ref~\cite{Autin:2002ms,Bouchez:2003fy,Mezzetto:2003ub} it is assumed that 
$2.9\cdot10^{18}$ $^6$He and $1.1\cdot10^{18}$ $^{18}$Ne 
decays per year can be achieved at a $\beta$-beam scenario 
with the acceleration $\gamma(^{18}Ne)=100$ and 
$\gamma(^{6}Ne)=60$ at the same time. If not stated
differently, we assume that the number of decays per year 
stays constant with varying $\gamma$. However, since this 
assumption is most likely not justified from technological 
considerations, we also describe effects 
where the number of decays varies with
$\gamma$ with the above values as normalization at
$\gamma(^{18}Ne)=100$ and $\gamma(^{6}Ne)=60$.

For all discussed $\beta$-beam setups, we choose a total 
running time of 8 years. This implies that the number of 
ion decays above are assumed to be reached by either the 
simultaneous operation of the neutrino and anti-neutrino 
beams, or their {\bf double} numbers by the successive 
operation of four years neutrinos and four years anti-neutrinos. 
We will also discuss deviations from equal neutrino and 
anti-neutrino running which could be achieved by 
successive operation option or by changing the ratio 
of stored ions. However, as default we use equal running 
times for neutrinos and anti-neutrinos, since the performance
for quantities highly correlated with $\deltacp$ is usually 
best for equal statistics in the neutrino and anti-neutrino 
channels, and the higher anti-neutrino flux turns out to 
compensate for the lower anti-neutrino cross section very well.

\subsection{Detector technologies}

A general requirement for any $\beta$-beam detector is to 
have good muon-electron separation capabilities and
to have an efficient neutral current rejection. At the same
time, the technology must be available and cost effective 
to allow in time a scaling to large detectors. For lower values 
of $\gamma$, certainly Water Cherenkov detectors (WC) fulfill these 
criteria~\cite{Zucchelli:2002sa,Mezzetto:2003ub,Burguet-Castell:2003vv,Donini:2004iv,Donini:2004hu,Burguet-Castell:2005pa}.
However, at higher $\gamma$ values, the lack of background 
discrimination in WCs becomes a huge problem and other 
detector types, such as calorimeters or TPCs (Time Projection 
Chambers) are more suitable~\cite{Burguet-Castell:2003vv,Ereditato:2004ru}.
The precise value of $\gamma$ where this turnover happens 
seems to be an unresolved question and quite different views can be 
found in the literature~\cite{Burguet-Castell:2003vv,Mezzetto:2003ub}. 
We will describe our own approach to this problem and we will discuss 
our findings in relationship to the existing literature. Our choice 
for large values of $\gamma$ is the so called 
``Totally Active Scintillator Detector'' (TASD) for reasons which 
will be discussed below. The two detector technologies used in this 
study, namely WC and TASD, are described in more detail in the 
following subsections.

\subsubsection{Water Cherenkov detector}
\label{sec:wc}

Water Cherenkov detectors are well suited to distinguish muon
neutrinos from electron neutrinos. However, background rejection
can be a problem in using a WC detector in combination with a 
$\beta$-beam. The main source of background to the muon neutrino 
appearance search will be the flavour-blind neutral current events which are 
mistaken for muon neutrino charged current events. 
The most critical neutral current 
events are those where one or several energetic pions are 
involved, which implies that there is basically no background
below the pion production threshold around $200\,\mathrm{MeV}$. 
Therefore, one solution would be to tune $\gamma$ to a low value 
where most of the neutrinos in the beam are below this 
threshold~\cite{Mezzetto:2003ub}. In that case there would be no 
energy information, since the Fermi-motion of the nucleons would 
induce an energy smearing of about $100\,\mathrm{MeV}$. This would 
reduce the $\beta$-beam to a mere counting experiment, which would 
have only a very limited physics reach~\cite{Burguet-Castell:2003vv}. 
Above the pion threshold, the feasibility of using a WC detector 
depends on the ability to correctly identify pions and to reject 
neutral current events. The pion identification works, in principle, 
by identifying its decay process and it seems to be possible up 
to some level. There are very different statements in the literature
how well this can
be done~\cite{Burguet-Castell:2003vv,Mezzetto:2003ub}. 
The different results can to a large extent be attributed to the 
different level of detail used in the detector simulation. Nonetheless
we will show  a direct comparison 
of the two simulations and our parameterization at a reference
scenario with $\gamma=150$ at a baseline of $L=300\,\mathrm{km}$ 
and an overall exposure equivalent to $5000\,\mathrm{kt}\,\mathrm{y}$,
where we call the one from
\Ref~\cite{Mezzetto:2003ub,Mezzetto:private,MezzettoTalkNNN:2005} 
``case~A'' and the one from
\Ref~\cite{Burguet-Castell:2005pa} 
``case~B''. As one can see from \Tab~\ref{tab:ab_comp}, the number 
of signal events is very similar in both cases, but the number of 
background events is very different. Moreover, the shape of the 
background spectra is very different and the background events are 
much more concentrated at low energies for case~B.

\begin{table}[t]
\begin{center}
\begin{tabular}{lcccccc} \hline
&\multicolumn{2}{c}{Case A}&\multicolumn{2}{c}{Case B}&\multicolumn{2}{c}{This work}\\
&$\nu$&$\bar\nu$&$\nu$&$\bar\nu$&$\nu$&$\bar\nu$\\
\hline
Signal&3128&3326&2956&3261&3129&3313\\
Background&514&588&176&123&170&123\\ \hline
\end{tabular}
\end{center}
\mycaption{\label{tab:ab_comp} The total signal and background rates in a 
Water Cherenkov detector at $L=300\,\mathrm{km}$ for an overall exposure 
of $5000\,\mathrm{kt}\,\mathrm{y}$ with a $\beta$-beam at $\gamma=150$ 
for neutrinos and anti-neutrinos. The oscillations parameters are the 
ones from \equ{standard_params} and $\sin^22\theta_{13}=0.1$. 
``case~A'' corresponds to \Ref~\cite{Mezzetto:2003ub,Mezzetto:private,MezzettoTalkNNN:2005} 
and ``case~B'' to \Ref~\cite{Burguet-Castell:2005pa}.}
\end{table}

The simulation of case~B is based on the Super-Kamiokande
Monte-Carlo~\cite{Gomez-Cadenas:2001eu} and it seems to be more 
detailed in its treatment of detector effects. For this reason
we use a parameterization which is, in total rates very close to 
case~B, as can also be seen from \Tab~\ref{tab:ab_comp}. Note, 
however, that even though the Monte-Carlo used for case~B 
has been well-tested in Super-Kamiokande, it is important to
keep in mind that such simulations rely on physical input such 
as cross sections, which are not very well known. Moreover, the
assumption that the response of a 20 times larger detector
is the same as that of Super-Kamiokande is implied there.

In order to describe the energy response of the detector in our
study, we divide the signal events into samples of quasi-elastic
events (QE) and inelastic events (IE). Only for the QE sample, it
is possible to accurately reconstruct the neutrino energy from
the charged lepton. For IE events, the reconstructed energy will
always lie below the true (incident) neutrino energy because
the hadronic component of the interaction cannot be seen by a
WC detector. Since the separation of those two event samples
is fraud with a large error, we will use the same technique as
described in \Ref~\cite{Huber:2002mx}. This means that the
total rates number of all $\mathrm{IE} \, +\, \mathrm{QE}$ events is
taken and in addition the spectrum of the QE event sample
is used to obtain spectral information. In order to avoid double
counting of events, the QE event sample is taken only with a
free normalization. In this approach, no particular assumption
about the event by event separation has to be made, because
this approach is purely statistical. In fact, the real experiment 
might even perform better since there actually could be some 
event by event separation. For the spectral analysis, we assume 
that the Fermi-motion is the main component of the resulting
energy resolution function. Therefore, a constant width of
$85\,\mathrm{MeV}$~\cite{Itow:2001ee} in a Gau\ss ian
energy resolution function is taken, in
order to describe the energy reconstruction of the QE sample.
For the background distribution, we make the assumption  
that every neutrino which interacts via neutral currents is
reconstructed with an energy distribution which is flat from 
zero up to the true neutrino energy. In this way we obtain a 
background which is peaked at low energies very similar to the 
one in case~B. We do not take into account any other background 
source, since in \Ref~\cite{Burguet-Castell:2005pa} it was 
shown that atmospheric neutrinos only give a very small contribution.

\begin{figure}[t]
\begin{center}
\includegraphics[width=0.5\textwidth]{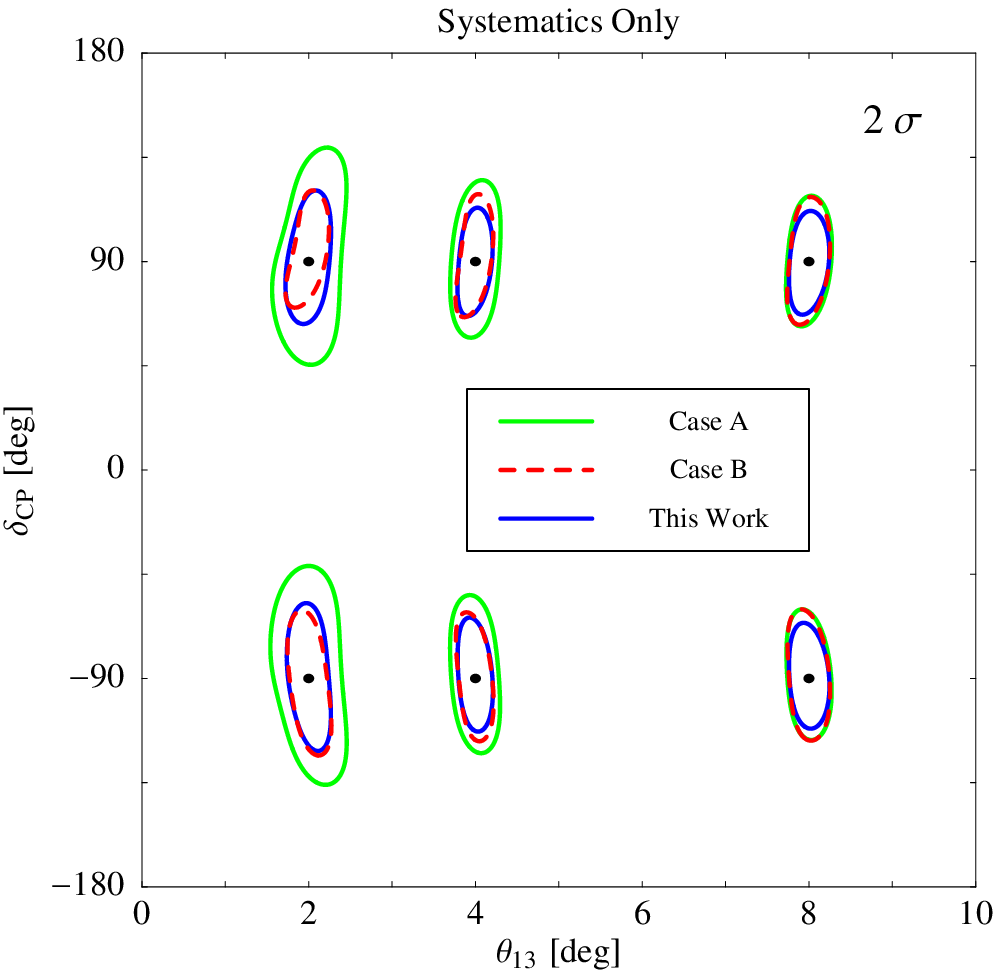}
\end{center}
\mycaption{\label{fig:ab_chi} The allowed regions in the 
$\theta_{13}$-$\deltacp$-plane at $2\sigma$ CL with only 
systematics taken into account. The solid green (bright
gray) curve corresponds to case~A, the dashed red (dark gray) 
curve to case~B, and the solid blue (black) curve to the scenario
used in this study. The black dots indicate the values of 
$\theta_{13}$ and $\delta$ used as input (true values)
and the other oscillation parameters are the ones from 
\equ{standard_params}.}
\end{figure}

\figu{ab_chi} shows a comparison of the physics output of the 
cases A, B and our parameterization. The allowed regions in the 
$\theta_{13}$-$\delta$ plane at a confidence level of 
$2\,\sigma$ for 2 d.o.f. are shown assuming that all the other 
parameters are exactly known and have the values as in 
\equ{standard_params}.
Case~B (dark gray/red curves) and our parameterization 
(black/blue curves) agree very well throughout the 
parameter space. The numbers used for our parameterization 
can be found in \Tab~\ref{tab:wc}.

In order to be able to use $\gamma$ values in the range 
from 50 to 500, the energy range is chosen as  
$0.2-3.0\,\mathrm{GeV}$ and is divided into bins 
of $100\,\mathrm{MeV}$ width, such that the total number 
of bins is 28. All efficiencies are constant with exception 
of the first bin where they are only $1/2$ of the value given 
in \Tab~\ref{tab:wc} to account for threshold effects. 
The numbers for the disappearance channel could be quite 
different (in fact, the efficiencies might be much higher), 
but we checked that the final results do not depend on 
this assumption. Therefore, because of simplicity, we take 
the same numbers as for the appearance channels. We also 
include systematic uncertainties on the normalization of
signal and background events as given in \Tab~\ref{tab:wc}, 
where all errors are assumed to be fully uncorrelated. 
This is a conservative approximation and does not affect
the result from the appearance measurement. The disappearance 
measurement, on the other hand, would require a total absolute 
error of less than 1\% to yield any information on 
$\theta_{13}$~\cite{Kozlov:2001jv,Minakata:2002jv,Huber:2003pm}. 
This number, however, seems very difficult to be reached, 
if at all, for any experiment which involves neutrino-nucleus
cross-sections at low energies. Our parameterization has been 
calibrated against case~B for \mbox{$\gamma=150$}. Therefore using
different values of $\gamma$ involves some extrapolation. 
Our parameterization should nevertheless be reliable from 
$\gamma$ 100 up to 350, and it should additionally reproduce 
the qualitative features of the $\gamma$-scaling within a 
range from 50 to 500.

\begin{table}[t]
\begin{center}
\begin{tabular}{lcccc} \hline
&\multicolumn{2}{c}{Appearance}&\multicolumn{2}{c}{Disappearance}\\
&$\nu$&$\bar\nu$&$\nu$&$\bar\nu$\\
\hline
Signal efficiency &0.55&0.75&0.55&0.75\\
Background rejection &0.003&0.0025&0.003&0.0025\\
Signal error &2.5\%&2.5\%&2.5\%&2.5\%\\
Background error &5\%&5\%&5\%&5\%\\ \hline
\end{tabular}
\end{center}
\mycaption{\label{tab:wc} The signal efficiencies and background
rejection respectively and the systematical errors for the various
signals and backgrounds used in our description of the performance
of a Water Cherenkov (WC) detector.}
\end{table}

\subsubsection{Totally active scintillator detector}
\label{sec:tasd}

For even larger values of $\gamma$, a detector which can also
measures the hadronic energy deposition is required. The reason
is that the fraction of inelastic events in the whole event
sample increases with increasing $\gamma$, because the neutrino
spectra are extended to higher energies. The techniques which
have traditionally been used for that purpose are tracking
calorimeters and TPCs (such as a large liquid Argon TPC as
described in \Ref~\cite{Ereditato:2004ru}). The latter technology
has certainly a great potential in neutrino physics, but given
the fact that background issues are not the primary concern,
we will discuss the more traditional and better understood
option of a tracking calorimeter. Basically, there exist three
different approaches:
\begin{itemize}
\item magnetized iron plates, interleaved with scintillator bars
\item low-Z material (such as particle board), interleaved with scintillator bars
\item all active detector made of liquid scintillator and plastic tubes
\end{itemize}
The big advantage of a (magnetized) iron calorimeter is usually
the ability to determine the charge of muons, but this is pointless
for a $\beta$-beam since there is no appearance of wrong sign muons
like at a neutrino factory. For the other options, the advantages and
disadvantages have been very carefully addressed in the preparation
of the NO$\nu$A proposal~\cite{Ayres:2004js}. We decided to use for
our study the same technology as in the NO$\nu$A proposal, which is
the so called ``Totally Active Scintillator Detector'' (TASD). The
totally active design provides a superior energy resolution and
background rejection at reasonable efficiencies. For our
parameterization, we follow closely the work done by the NO$\nu$A
collaboration, the only problem being that all studies have been
done for $\nu_e$ appearance, whereas we look for $\nu_\mu$
appearance. The latter should be much easier because the muon track
is much more difficult to be confused with a neutral current event. 
Therefore, our parameterization is on the conservative side, which 
does not affect our conclusions since the TASD very effectively 
rejects backgrounds. The numbers we use for efficiencies and 
systematical errors are given in \Tab~\ref{tab:tasd} and are taken 
from~\Refs~\cite{Ayres:2004js,SIM-42,SIM-48,private_Litchfield}.

The energy window reaches from $0.5\,\mathrm{GeV}$ up to the endpoint
of the neutrino spectrum and is divided into 20 bins.  The energy
resolution is given by a Gau\ss ian with a width of 3\% $\sqrt{E}$ for
muon neutrinos and 6\% $\sqrt{E}$ for electron neutrinos. The
background is assumed to have the same shape as the signal. But note
that the shape of the background is not much of an issue in the case
of a TASD detector since the background is very small. We checked that
a background of the same total magnitude which is distributed like
$E_\nu^{-1}$ gives basically the same results.

\begin{table}[t]
\begin{center}
\begin{tabular}{lcccc} \hline
&\multicolumn{2}{c}{Appearance}&\multicolumn{2}{c}{Disappearance}\\
&$\nu$&$\bar\nu$&$\nu$&$\bar\nu$\\
\hline
Signal efficiency&0.8&0.8&0.2&0.2\\
Background rejection&0.001&0.001&0.001&0.001\\
Signal error&2.5\%&2.5\%&2.5\%&2.5\%\\
Background error&5\%&5\%&5\%&5\%\\ \hline
\end{tabular}
\end{center}
\mycaption{\label{tab:tasd} The signal efficiencies and background rejection
respectively and the systematical
errors for the various signals and backgrounds used in our description of a TASD.}
\end{table}
%

\subsection{Experiment configurations and event rates}
\label{sec:other}

In order to compare the different detector options we still need 
to define the detector size and the luminosity of the beam. The 
fiducial volume of a WC detector seems to lie naturally of the 
order of $1\,\mathrm{Mt}$, since this suits also other applications
as proton decay searches and there exists various proposals of
this type, see \eg\ \Ref~\cite{Jung:1999jq}. For definiteness, 
we assume our WC detector of this type, but with a somewhat 
more affordable fiducial volume of $500\,\mathrm{kt}$. For the 
TASD there exists currently only the NO$\nu$A proposal for a 
$30\,\mathrm{kt}$ detector, and $50\,\mathrm{kt}$'s seem to be 
feasible. We will use therefore the latter (larger) value as the
detector mass of the TASD within this study. Within the usual 
uncertainties, these two detectors should also have a comparable
price. These two choices lead to the typical signal and background 
event rates shown in \Tab~\ref{tab:events} for the true parameters
of \equ{standard_params} and $\sin^22\theta_{13}=0.1$. Note, however,
that these numbers are calculated under the assumption of a constant 
number of isotope decays. This will of course change when we include
the $\gamma$-scaling in subsequent sections.

\begin{table}[t]
\begin{center}
\begin{tabular}{lrrr} \hline
Detector type&WC&TASD&TASD\\ \hline
$m\,[\mathrm{kt}]$&500&50&50 \\
$\gamma$&200&500&1000\\
$L\,[\mathrm{km}]$&520&650&1000\\ \hline
$\nu$ signal&1983&2807&7416\\
$\nu$ background&105&31&95\\ \hline
\end{tabular}
\end{center}
\mycaption{\label{tab:events} The number of signal/background events
for different combinations of the chosen detector type
and values of $\gamma$. These numbers are calculated for constant number of isotope
decays per year with varying $\gamma$.
The oscillation parameters are the same as in \equ{standard_params} with
$\sin^22\theta_{13}=0.1$.}
\end{table}
%

\section{Optimization of $\stheta$ sensitivity}
\label{sec:optimals13}

In this section, we focus on the optimization of detecting a finite value
and/or measuring $\stheta$. For that, we introduce performance indicators
for the sensitivity to $\stheta$ and discuss the principle degrees of 
freedom for the optimization. Furthermore, we illustrate the optimization 
in the most relevant directions of the parameter space and choose specific 
setups. Eventually, we compare the performance to other established
techniques, such as neutrino factories or superbeam upgrades.

\subsection{Degrees of freedom and performance indicators}

The optimization of a $\beta$-beam experiment involves a number of issues:
\bi
 \item
  What is the optimal $\gamma$? Obviously, the detector technology
  is a major issue in this optimization. In addition, external constraints
  may cause the number of decays per year not to stay constant with
  increasing $\gamma$.
 \item
  What is the optimal baseline for a given $\gamma$?
 \item
  How long should one run in the $\nu$ and $\bar{\nu}$ running mode?
  At the same or different $\gamma$'s?
  Can one run these modes simultaneously?
 \item What isotopes should one use? How many decays per year are realistic?
\ei

We will discuss some of these issues in greater detail below, while
we will make reasonable assumptions in other cases. Let us first
repeat the main assumptions which have already been discussed: We assume
neutrino and anti-neutrino beams from the decay of $^6$He and $^{18}$Ne
isotopes with the reference numbers for the decays per year at $\gamma(^6He) =60$
and $\gamma(^{18}Ne) = 100$ as in \Ref~\cite{Autin:2002ms,Bouchez:2003fy,Mezzetto:2003ub}.
In general, we assume a successive operation with neutrinos and anti-neutrino running,
since we will allow a variation of the neutrino versus antineutrino running time
in some cases.
Furthermore, we fix $\gamma(^6He) = \gamma(^{18}Ne)$,
since there is no obvious gain in increasing one of the two $\gamma$'s
(\cf, \Ref~\cite{Burguet-Castell:2005pa}). We also use a constant,
$\gamma$ independent detector mass for an assumed Water Cherenkov (WC)
or a Totally Active Scintillator Detector (TASD). The WC detector will
be used only below $\gamma \sim 500$, since for higher energies the WC
detector will be dominated by non-QE events and the background
parameterization is rather unclear.

Somewhat more complicated is the issue of the $\gamma$-dependence of
the beam. Initially we will assume that the number of decays per year
does not depend on $\gamma$, but this is definitively not realistic.
We will therefore discuss in detail the impact of a scaling with
$\gamma$ later.

We use two different performance indicators for $\stheta$.
We define the $\stheta$ sensitivity as the largest value of
$\stheta$ which fits $\stheta=0$, \ie,  the $\stheta$ sensitivity
tests the hypothesis $\stheta=0$. It corresponds to the new exclusion
limit if an experiment does not observe a signal. Since the simulated
rate vector is computed for $\stheta=0$, this sensitivity does not
depend on the true (simulated) values of $\stheta$, $\deltacp$, or the
mass hierarchy (\cf, \App~C of \Ref~\cite{Huber:2004ug}). However, there
are strong correlations and degeneracies in the fit rate vector because
any combination of parameters which fits $\stheta=0$ destroys the
$\stheta$ sensitivity. In particular, we compute the statistics and
systematics $\stheta$ sensitivity for fixed $\deltacp=0$ and the other oscillation
parameters fixed to their simulated values. The correlation
with $\deltacp$ and the other oscillation parameters will then be included by
the projection of the fit manifold onto the $\stheta$-axis as the correlation bar of our
figures, where only the best-fit solution is used. Any disconnected solution
at the chosen confidence level which fits $\stheta=0$ is treated as
degeneracy, such as a $(\deltacp,\theta_{13})$~\cite{Burguet-Castell:2001ez}
or mass hierarchy~\cite{Minakata:2001qm} degeneracy. Thus, we treat connected
degenerate solutions (with the best-fit manifold) as correlations, and disconnected
degenerate solutions as degeneracies.

In addition to the $\stheta$ sensitivity, we show the $\stheta$ discovery
reach in some cases, which tests the hypothesis of observing a signal
for a given set of true values. Thus, the $\stheta$ discovery reach
strongly depends on the true values of $\stheta$, $\deltacp$, and the
mass hierarchy. However, there are almost no correlations or degeneracies,
since the fit rate vector is computed for $\stheta=0$. In principle,
the risk-minimized (with respect to the possible true values) $\stheta$
discovery reach is comparable to the $\stheta$ sensitivity -- though
the problem is not exactly symmetric.

\subsection{Performance as function of $\gamma$ and baseline optimization}

\begin{figure}[t]
\begin{center}
\includegraphics[width=\textwidth]{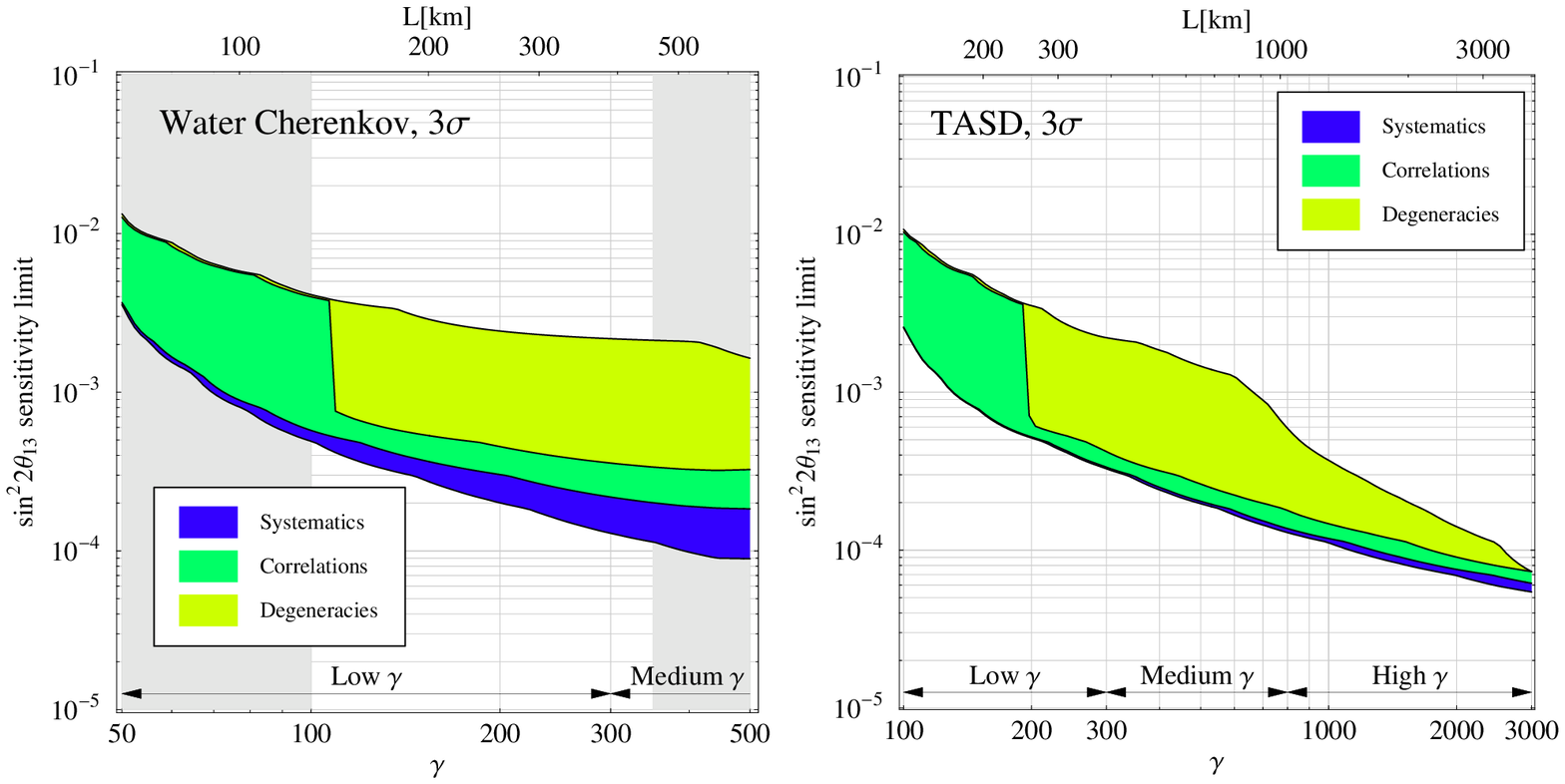}
\end{center}
\mycaption{\label{fig:detscaling} The $\stheta$ sensitivity limit as
function of $\gamma$ for a  constant number of decays per year
(as function of $\gamma$) and $L=1.3 \, \gamma$ at the $3 \sigma$ confidence level. The
plots are for the Water Cherenkov detector (WC, left) and Totally
Active Scintillator Detector (TASD, right). The final sensitivity
limits are obtained as the upper edges of the curves after successively
switching on systematics, correlations, and degeneracies.}
\end{figure}

A first important optimization question concerns the optimal value
of $\gamma$. Naively this question seems to be trivial for a fixed
number of decays per year - the higher the $\gamma$, the better.
However, there is a strong
dependence on detector technology, since non-QE events will start to
dominate a WC detector for higher energies, and the efficiency of a TASD
(or iron calorimeter) is very low at low energies because of too short
tracks compared to the positional information of the detector.
We show therefore in \figu{detscaling} the $\stheta$ sensitivity for
the WC detector (left) and the TASD (right) as function of $\gamma$,
where the parameterization of the WC detector is most reliable in the
unshaded region. For the $L/\gamma$, we assume 1.3 for both detectors.
Note, however, that we will find that
$L/\gamma=1.3$ is not optimal for the Water Cherenkov detector
if one includes correlations and degeneracies.
For the TASD, the rule ``the higher the
$\gamma$, the better'' obviously applies if the stored ion decays are constant
in $\gamma$, since the muon tracks are in
average fully contained in the shown range. For the WC detector, this
rule also applies in the unshaded region, but it is likely that
background domination and non-QE events will affect the performance
for higher $\gamma$. In addition, the dependence on $\gamma$ is much more
shallow in the considered range, and the impact of systematics increases
with increasing $\gamma$. Note that though the Water Cherenkov detector
has the better systematics only performance at the upper end of the unshaded region
(left panel), both detectors perform very similar after including correlations and degeneracies in
this range.

\begin{figure}[tp]
\begin{center}
\includegraphics[width=\textwidth]{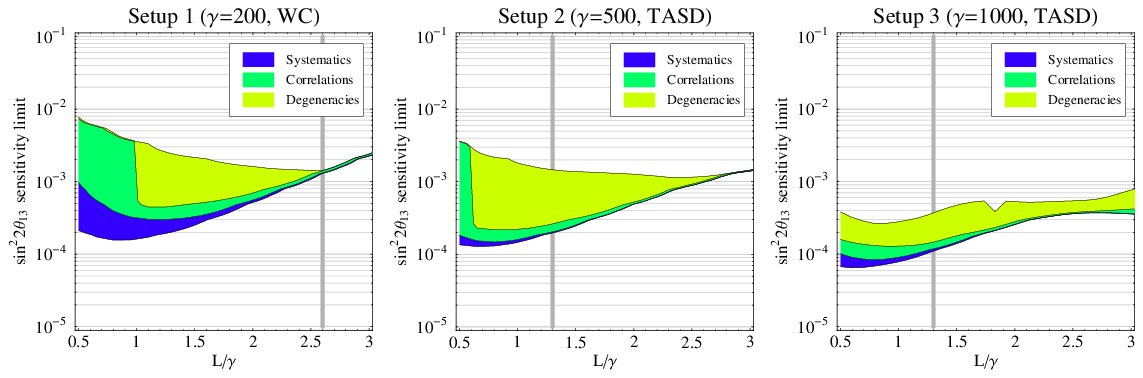}
\end{center}
\mycaption{\label{fig:ldepn0} The $\stheta$ sensitivity limits at
the $3\sigma$ confidence level for
the different setups as function of the ratio $L/\gamma$ for the number of
decays per year fixed. Note that $\gamma$ in each individual plot
is fixed to the value given in the respective figure caption.
The final sensitivity
limits are obtained as the upper edges of the curves after successively
switching on systematics, correlations, and degeneracies.}
\end{figure}

The baseline dependence of the $\stheta$ sensitivity on $L/\gamma$
is shown in \figu{ldepn0} for different choices of (fixed) $\gamma$
as given in the plot labels. The optimal $\stheta$ sensitivity can in all cases
be achieved from a statistical and systematical point of view for
$L/\gamma \sim 0.8 - 1.3$, where the appearance events only come from the
first oscillation maximum. For higher values of $L/\gamma$ the statistics 
$\stheta$ sensitivity decreases due to the $1/L^2$ dependence of the flux
although the actual position of the first oscillation maximum, taking into account the
average neutrino energy, would be at $L/\gamma \sim 2.1$.
However, longer baselines, where appearance events from the second oscillation maximum
enter the energy window of the analysis, have altogether a better potential to resolve
correlations and degeneracies. The combined effect of the first and
second oscillation maximum together leads to a better determination of the
oscillation pattern and larger matter effects, though the statistic
limit becomes worse.
For the WC detector, we therefore choose $L/\gamma = 2.6$, where
the impact of correlations and degeneracies is marginal and the overall
performance is significantly improved. We demonstrate in
Appendix~A in detail where this better performance comes
from and why it is only visible after the inclusion of correlations
and degeneracies. Furthermore, we choose $L/\gamma=1.3$
for the TASD, since a larger $L/\gamma$ hardly improves the
performance. In addition, we have tested that if the number of
decays per year significantly drops with increasing $\gamma$,
the minimum of Setup 3 is shifted towards longer baselines. Note
that in all cases the flatness in $L/\gamma$ implies that the precise
baseline is not so important for the overall $\stheta$ sensitivity.
In addition, the choice of these values for $L/\gamma$ cannot be entirely
based upon this figure and will be later justified in the context of
different performance indicators.

\begin{figure}[t]
\begin{center}
\includegraphics[width=0.5\textwidth]{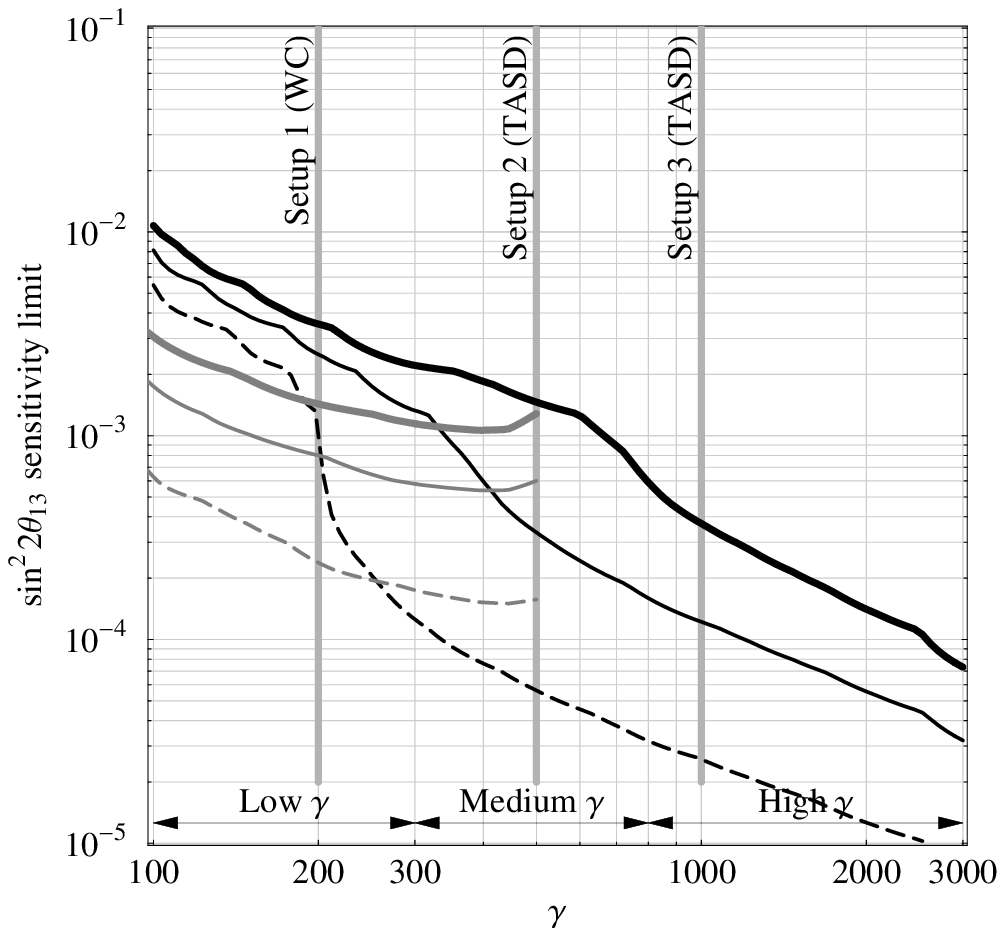}
\end{center}
\mycaption{\label{fig:detcomp} The final $\stheta$ sensitivity limit
(including systematics, correlations, and degeneracies) as function
of $\gamma$ for a fixed number of decays per year. The curves correspond
to the Totally Active Scintillator Detector ($L=1.3 \, \gamma$,
TASD, black curves) and the Water Cherenkov detector ($L=2.6 \, \gamma$,
WC, gray curves) at the $1 \sigma$ (dashed), $2 \sigma$ (thin solid),
and $3 \sigma$ (thick solid) confidence levels. The vertical lines
correspond to our low-, medium-, and high-$\gamma$ setups. Note that
the WC parameterization is only reliable for
$100 \lesssim \gamma \lesssim 350$.}
\end{figure}

Let us now directly compare the two detector technologies in \figu{detcomp},
where the gray curves correspond to the WC detector and the black curves
to the TASD. For small $\gamma$, the WC detector obviously has the best performance,
whereas for higher $\gamma$, the TASD is the way to go for.
The crossing point between the two technologies depends on the confidence
level and lies somewhere in the interval $250 \lesssim \gamma \lesssim 500$.
Note, however, that the TASD performance is dominated by correlations
and degeneracies (\cf, \figu{detscaling}),
which means that its discovery potential will certainly be better for
$\gamma =500$ at all confidence levels. In addition, the parameterization
of the WC detector is not very reliable in this region anymore.

In order to discuss some effects in greater detail we use well defined
setups/representatives in order to evaluate the requirements for a
$\beta$-beam. Comparing with \figu{detcomp}, there are three
interesting (approximate) ranges:
\begin{enumerate}
 \item Low $\gamma \lesssim 300$: This range can be probed with
 relatively ``small'' accelerators,  such as of SPS size, and
 WC detectors. The physics potential could compete with superbeam
 upgrades, such as an upgrade of T2K to a Hyperkamiokande detector,
 or an intermediate step towards a neutrino factory.
 \item Medium $300 \lesssim \gamma \lesssim 800$: In this case,
 larger accelerator rings are required, for example of Tevatron size.
 The detector technology could be WC or TASD. The
 physics goal could be to compete with superbeam upgrades or
 even small neutrino factories.
 \item Large $\gamma \gtrsim 800$: Very large accelerators of
 the size of the LHC are required\footnote{Note, however, that even
 though the LHC would have enough energy, it certainly does not
 have enough RF acceleration power, \ie, a new huge accelerator would be
 required in this case.}.
 TASD or iron calorimeters are possible choices for the detector
 technologies. In this case, the goal is clearly competition with
 neutrino factories for all of the relevant measurements.
\end{enumerate}
These $\gamma$-ranges are shown in many figures of this and the
next chapter. Obviously, the requirements for all of these
ranges are somewhat different, which means that it will not only
be sufficient to compare $\beta$-beam with $\beta$-beam, but it
will also be necessary to compare individual $\beta$-beams from
each range with its competitors. For this purpose, we define three
setups from these three ranges (\cf, \figu{detcomp}):
\begin{itemize}
\item
 {\bf Setup 1 ($\gamma=200$, WC):} In this case, the WC
 detector representation is well-established and $\gamma$ should
 not be too high for SPL-sized accelerators.
\item
 {\bf Setup 2 ($\gamma=500$, TASD):} This setup represents the lowest
 $\gamma$ where a TASD will likely perform better that a WC
 detector at all confidence levels.
\item
 {\bf Setup 3 ($\gamma=1000$, TASD):} This representative corresponds to
 a very sophisticated option close the upper limit of what is doable.
\end{itemize}

\subsection{Isotope decay scaling}

We have already mentioned that the number of decays per year is
most likely not constant in $\gamma$. In order to include this
effect, we use the following power law parameterization, which should be
justified for a certain $\gamma$-range, to describe this
scaling with $\gamma$ ($i=1$ for $^{18}Ne$: neutrinos, $i=2$
for $^{6}He$: anti-neutrinos):
 \begin{equation}
  N^i = N_0^i \cdot \left( \frac{\gamma_0^i}{\gamma} \right)^n
 \end{equation}
 Here $N_0^i$ is determined by our reference point at
 $\gamma_0^1=100$, $\gamma_0^2=60$. We can now discuss different
 cases for $n$, which leads to different optimization strategies:
 \bi
 \item
  $n=0$: The number of decays per year is fixed. This implies that
  the accelerator and storage ring has to scale appropriately with
  $\gamma$ in a non-trivial manner.
 \item
  $0<n<1$: This seems to be the most likely range of realistic cases.
  The number of decays per year becomes constrained with increasing
  $\gamma$ by the geometry of the accelerator and decay ring and
  $\gamma$ increased lifetime of the isotopes in the laboratory system.
 \item
  $n\sim 1$: This case corresponds to a fixed setup constraining the
  performance. Given the scaling of the baseline, the number of
  events stays approximately constant as function of $\gamma$.
  A realistic constraint for the SPS would be, for example, $n \sim 1$
  from the number of merges in the decay ring and the number of ions
  per bunch \cite{Lindroostalk}.
 \item
  $n>1$: In this case, it clearly does not make sense to go to higher
  $\gamma$'s, since the event rate decreases with $\gamma$ if we stay
  in the oscillation maximum
 \item
  $n<0$: The number of decays per year increases with $\gamma$. This
  hypothetical (but technologically unlikely) possibility requires
  that the accelerator and decay ring over-proportionally scale with
  $\gamma$.
\ei
We further on consider the range $0 \lesssim n \lesssim 1$ to be
realistic. However, it is conceivable that, for a given setup, the
performance will  scale with $n\simeq 0$ in the beginning, and
change into an $n\simeq 1$ scaling in the saturation regime.

\begin{figure}[t]
\begin{center}
\includegraphics[width=0.5\textwidth]{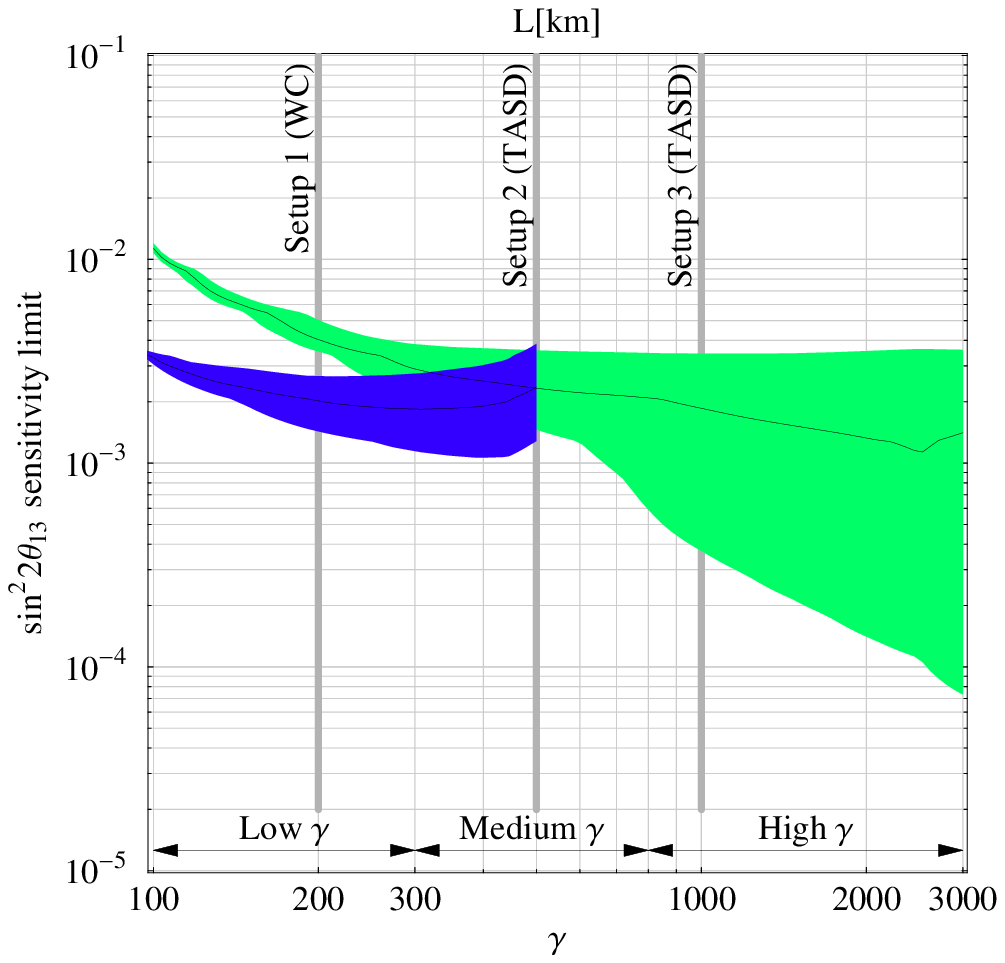}
\end{center}
\mycaption{\label{fig:ndepgamma} The final $\stheta$ sensitivity
limit (including systematics, correlations, and degeneracies) as
function of $\gamma$ for $L=1.3 \, \gamma$ (TASD) and
$L=2.6 \, \gamma$ (WC). The bands correspond to the Totally Active
Scintillator Detector (TASD, green/light shaded region) and the
Water Cherenkov detector (WC, blue/dark shaded region) at the
$3 \sigma$ confidence level, where the isotope luminosity
exponent $n$ is varied from $0$ (lower
ends of the bands) over $0.5$ (black curves) to $1$ (upper end
of the bands). The vertical lines correspond to our low-,
medium-, and high-$\gamma$ setups.}
\end{figure}

We show in \figu{ndepgamma} the $\gamma$-scaling of the final $\stheta$
sensitivity (including correlations and degeneracies) for the different
detector technologies and  $0 \lesssim n \lesssim 1$ (bands), where
the solid lines correspond to $n = 0.5$. For $n \sim 1 $ (upper ends
of bands), the performance is rather flat in $\gamma$. This simply
means that the increase in cross section is compensated by the
$\gamma$-scaling. For $n \sim 0$ (lower end of bands), the $\stheta$
sensitivity scales as discussed above. It is interesting to observe
that Setups~1 and~2 are much less affected by a decreasing number of
decays than Setup~3, which over-proportionally looses sensitivity for
$n>0$. Thus, for Setup~3, it is crucial that the accelerator and
storage design allow enough decays per year, whereas for Setups~1
and~2 a certain loss in the number of decays results only in about a
factor of two weaker sensitivity limit. This means that, given a
specific accelerator (such as the LHC), it makes only sense to discuss
very high $\gamma$ setups if it is guaranteed that enough ions can be
stored. Note that the dependence on $n$ is therefore an important
constraints for the $\gamma$ optimization of $\beta$-beams, since,
for the TASD, it constrains the rule ``the larger $\gamma$, the better''.

\subsection{Comparison with other technologies}
\label{sec:comparetech}

Next we compare the $\stheta$ sensitivity of $\beta$-beams to
neutrino factories and superbeam upgrades. For the neutrino factory,
we use the representative \NuFactII\ from \Ref~\cite{Huber:2002mx}
which uses only $\nu_\mu$ appearance and disappearance
in two different baseline configurations $L=3 \, 000 \, \mathrm{km}$ 
and $L=7 \, 500 \, \mathrm{km}$ (NF@3000km and NF@7500km).
This assumes $4 \, \mathrm{MW}$ target power (corresponding to about
$1.06 \cdot 10^{21}$ useful muon decays per year), a
$50 \, \mathrm{kt}$ fiducial volume magnetized iron calorimeter, 
and $50 \, \mathrm{GeV}$ neutrinos. The longer baseline corresponds 
to the ``magic baseline'', where correlations and degeneracies are 
resolved, but no $\deltacp$ measurement is possible~\cite{Huber:2003ak}. 
The $\stheta$ sensitivity and mass hierarchy sensitivity of the 
neutrino factory at the magic baseline is therefore very competitive. 
We include this option because we want to compare optimized setups
(for particular purposes) with optimized setups later, \ie, the 
optimal $\beta$-beam for the mass hierarchy measurement may also 
have a different baseline that the one for the CP violation measurement.
For the superbeam upgrade, we choose T2HK from \Refs~\cite{Itow:2001ee} 
simulated in \Ref~\cite{Huber:2002mx}, but we reduce the fiducial mass 
to $500 \, \mathrm{kt}$ and use the same detector as in this study in 
order to be comparable to our WC detector for Setup~1. This superbeam 
upgrade also assumes a target power of $4 \, \mathrm{MW}$ and we call
it therefore \JHFHK . The different setups are summarized in \tabl{setups}.

\begin{table}[t]
\begin{center}
\begin{tabular}{lrrrlrrc}
\hline
Label & $\gamma$ & L/km & $\langle E_{\nu} \rangle $/GeV& Detector & $m_{\mathrm{Det}}$/kt & $t_{\mathrm{run}}$/yr  & \Ref \\
 & & & & & & $(\nu,\bar{\nu})$ & \\
\hline
Setup~1 & 200 & 520 & 0.75 & Water Cherenkov & 500 & (4,4) &  \\
Setup~2 & 500 & 650 & 1.9 & TASD & 50 & (4,4) & \\
Setup~3 & 1000 & 1300 & 3.8 & TASD & 50 & (4,4) &  \\
\hline
\JHFHK\ & n/a & 295 & 0.76 & Water Cherenkov & 500 & (2,6) & \cite{Huber:2002mx} \\
NF@3000km & n/a & 3000 & 33 & Magn. iron calor. & 50 & (4,4) & \cite{Huber:2002mx} \\
NF@7500km & n/a & 7500 & 33 & Magn. iron calor. & 50 & (4,4) & \cite{Huber:2003ak} \\
\hline
\end{tabular}
\end{center}
\mycaption{\label{tab:setups} The experiment representatives used in
this study, where all detector masses are fiducial masses. Note that we
adjusted the detector mass of \JHFHK\ compared to \Ref~\cite{Huber:2002mx}
and use the identical detector to Setup~1.}
\end{table}

\begin{figure}[tp]
\begin{center}
\includegraphics[width=0.6\textwidth]{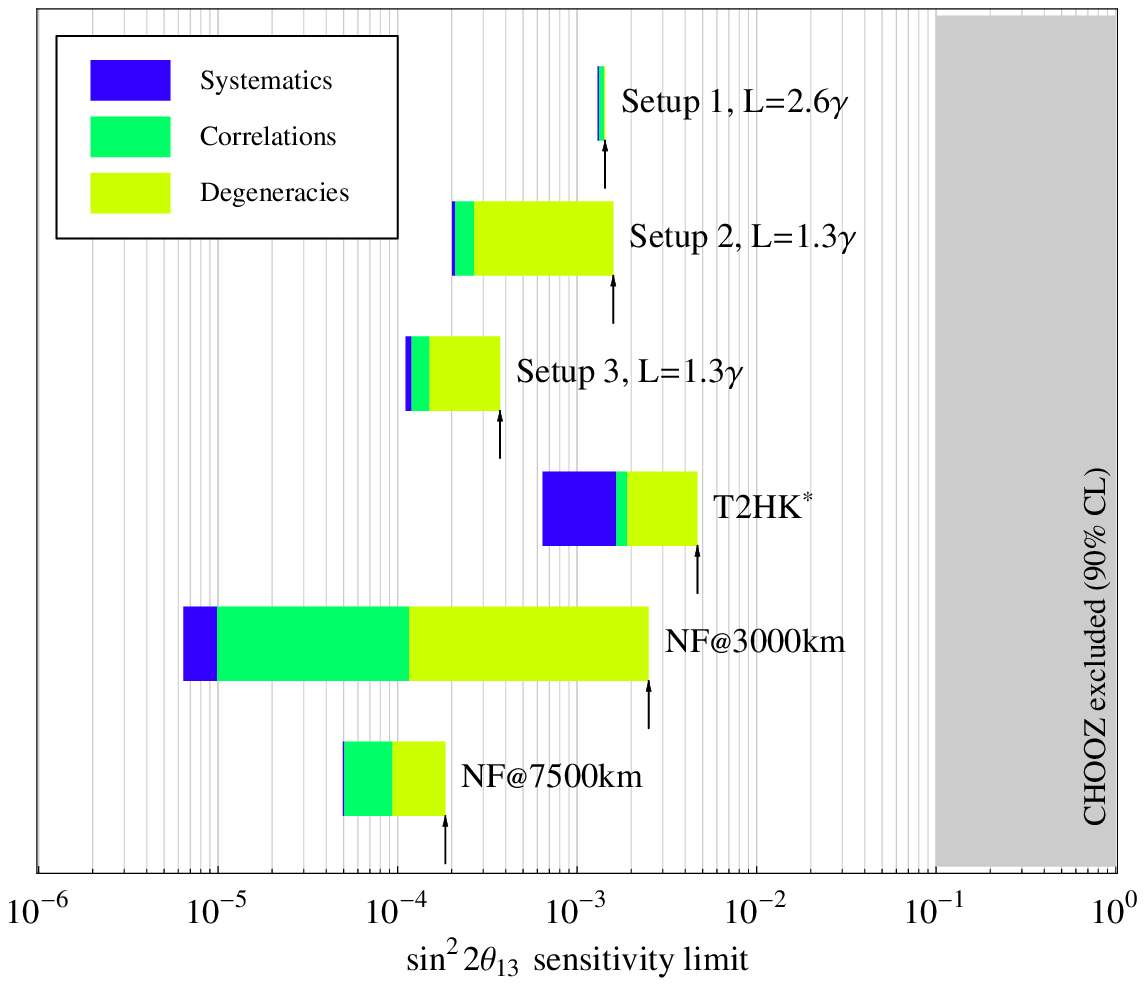}
\end{center}
\mycaption{\label{fig:barscomp} The $\stheta$ sensitivity limits for the
different setups and other representatives. Here $n=0$ (decays per year
fixed) and the $3 \sigma$ confidence level are chosen. The final sensitivity
limits are obtained as the right edges of the bars after successively
switching on systematics, correlations, and degeneracies.}
\end{figure}

\figu{barscomp} shows the $\stheta$ sensitivity for these setups.
First, it is interesting to observe that all of the $\beta$-beam
representatives are competitive to the \JHFHK\ setup in all cases.
The neutrino factory at $L=3 \, 000 \, \mathrm{km}$ can, for the
chosen parameter values, not resolve the $(\delta,\theta_{13})$-degeneracy
at the $3 \sigma$ confidence level\footnote{This behaviour could in
  principle change once additional information like $\nu_\tau$
  appearance is available}, and therefore the final $\stheta$
sensitivity is much worse than that of the $\beta$-beams. The neutrino
factory at $L=7 \, 500 \, \mathrm{km}$ has the best overall performance
because it is hardly affected by the correlation with $\deltacp$.
Note that Setup~1 has a better final $\stheta$-sensitivity than
Setup~2 because of the choice of $L/\gamma=2.6$. Though the statistics
limit is worse for this choice, the correlation and degeneracy bars become very
small and lead to a better final sensitivity than for $L/\gamma=1.3$.

\begin{figure}[t]
\begin{center}
\includegraphics[width=\textwidth]{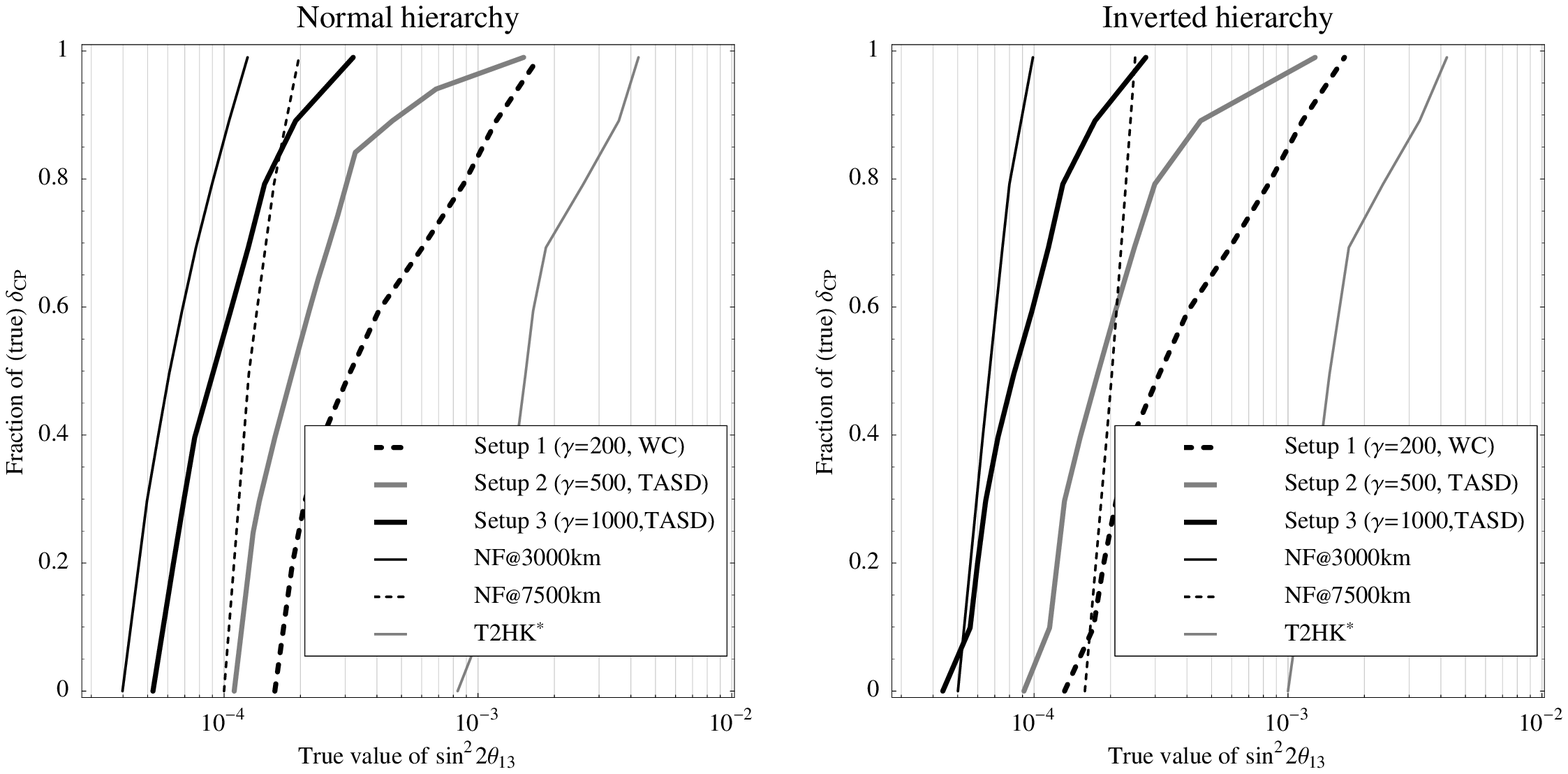}
\end{center}
\mycaption{\label{fig:t13cpf} The $\stheta$ discovery reach (including
systematics and correlations) for different setups as function of the 
true values of $\stheta$ and $\deltacp$ ($3 \sigma$ confidence level),
where the left plot is for the normal hierarchy and the right plot for 
the inverted hierarchy. The values of $\deltacp$ are ``stacked'' to the 
fraction of $\deltacp$, \ie, $\stheta$ will be discovered for a certain 
fraction of all possible values of $\deltacp$. For a uniform probability 
contribution in $\deltacp$, the fraction of $\deltacp$ directly 
corresponds to the probability to discover $\stheta$.}
\end{figure}

In order to complete the picture, we show in \figu{t13cpf} the $\stheta$
discovery reach as function of the true values of $\stheta$, $\deltacp$
(stacked to the ``CP fraction''), and the mass hierarchy. Though the
$\stheta$ sensitivity of Setup~1 is slightly better than one of
Setup~2, one can clearly see that there is a hierarchy in the discovery reach: The choice of
Setups~1,~2 and~3 implies that their differences correspond to approximately equal
improvements in terms of fraction of the parameter space.
The large impact of correlations on the $\stheta$ sensitivity of Setup~2 (\cf,
\figu{barscomp}), which mainly comes from the correlation with $\deltacp$,
implies that the discovery reach is in many cases of $\deltacp$ better
than the one of Setup~1.
Note that the relative position of the Setup~1 curve would not significantly change
with a different choice of $L/\gamma=1.3$ because in this case the
shape of the curves would be very similar, but the systematics limit of
Setup~2 is slighlty better (\cf, \figu{detscaling}).
For a normal hierarchy, the discovery reach of
Setup~3 covers considerably less parameter space than that one of a
neutrino factory at $L=3 \, 000 \, \mathrm{km}$. For the inverted
hierarchy, the matter effect enhancement of the lower neutrino factory
anti-neutrino rate (instead of the higher neutrino rate) leads to a
relatively degraded reach for the neutrino factory baselines, whereas
the event numbers of the $\beta$-beams are rather similar for neutrinos
and anti-neutrinos. However, the neutrino factory at
$3 \, 000 \, \mathrm{km}$ covers the most parameter space in both cases,
and the superbeam upgrade \JHFHK\ by far the least. Note that the neutrino
factory behavior for the $\stheta$ sensitivity and $\stheta$ discovery
reach is completely different. For the $\stheta$ sensitivity, very few
particular combinations of parameters prevent a strong $\stheta$ limit,
whereas for the $\stheta$ discovery reach, correlations and degeneracies
are of secondary importance. A neutrino factory is therefore clearly a
$\stheta$ discovery machine. Note that the discovery reach fit rate
vector is computed for (fixed) $\stheta=0$, which means there is only
a substantial impact of correlations if the solar appearance
term contributes significantly to the appearance
rate.\footnote{It turns out that these correlations have some impact on
the discovery reach of Setup~1 and the neutrino factory baselines,
especially $L=3 \, 000 \, \mathrm{km}$, since both, having a long
enough baseline and being far off the matter density resonance (in
energy) increase the relative importance of the solar appearance term.}
Since the solar appearance term does not depend on the mass hierarchy,
there is no contribution of the $\mathrm{sgn}(\ldm)$-degeneracy.

In summary, all of the discussed $\beta$-beam options could be 
interesting alternatives to a superbeam upgrade or intermediate 
options towards a neutrino factory. Especially Setup~1, which 
uses the identical detector needed for other applications, might be 
an interesting transition candidate. Since the $\beta$-beams have 
a clean composition of electron neutrinos, they are not affected 
by an intrinsic contamination of muon events and are therefore not
systematics limited close to $\stheta \sim 10^{-3}$ such as superbeams 
are. We will discuss the sensitivity to the mass hierarchy and CP 
violation in the next section in order to get a broader perspective 
of the problem, and to evaluate if a $\beta$-beam
could, in principle, replace a neutrino factory.

\section{Sensitivity to mass hierarchy and CP violation}
\label{sec:optimalHCP}

Beyond detecting a finite value of $\stheta$, the mass hierarchy and
CP violation sensitivities are the two most interesting quantities to
be measured by the discussed experiments. We will first introduce
performance indicators for these quantities. Then we will discuss
optimization aspects for the mass hierarchy and CP violation
determination. Finally, we will compare the performances of the $\beta$-beam
setups with other experiments.

\subsection{Performance indicators}

We define sensitivity to a particular mass hierarchy (normal or inverted)
if the wrong-sign solution can be excluded at a certain
confidence level. This implies that a $\mathrm{sgn}(\ldm)$-degeneracy~\cite{Minakata:2001qm}
or mixed degeneracy~\cite{Barger:2001yr} will destroy the mass hierarchy sensitivity.
The mass hierarchy sensitivity does not only depend on the simulated hierarchy, but also
on the true values of $\stheta$ and $\deltacp$. Since it is not possible to show the
full parameter space for mass hierarchy sensitivity simultaneously, we use in many
cases the sensitivity for the true $\deltacp=0$. In other cases, we show the mass
hierarchy sensitivity as function of $\stheta$ and $\deltacp$, where we stack the
true values of $\deltacp$ to the ``Fraction of $\deltacp$'' (CP fraction) similar to
\figu{t13cpf}. A mass hierarchy sensitivity
for a CP fraction $1$ corresponds to mass hierarchy sensitivity for any values of $\deltacp$
(worst case in $\deltacp$), and a mass hierarchy sensitivity for a CP fraction $\rightarrow 0$
to mass hierarchy sensitivity for the best case $\deltacp$. A CP fraction $0.5$ refers to
the ``typical'' value of $\deltacp$, \ie, 50\% of all cases of
$\deltacp$. Note that the CP fraction is a probabilistic measure in the sense
that only one of these values can be realized by nature. Assuming a uniform
distribution of $\deltacp$, it directly corresponds to the
probability to discover the mass hierarchy for a chosen $\stheta$.

We define sensitivity to CP violation if the CP conserving solutions $\deltacp=0$ and
$\deltacp=\pi$ can be excluded at a certain confidence level. This implies that any
degenerate solution overlapping a CP conserving solution destroys the
sensitivity to CP violation. Note that this sensitivity clearly differs from
the parameter sensitivity to a specific parameter value of $\deltacp$, which
includes the sensitivity to the special value $\deltacp=0$. In some cases, we 
only show the parameter sensitivity to maximal CP violation $\deltacp=\pi/2$ or 
$\deltacp=3\pi/2$. However, since any value of $\deltacp \neq \{0, \pi \}$ 
violates CP, we also show the parameter sensitivity to any
CP violation in other cases, \ie, the sensitivity to CP violation as function of
the true values of $\stheta$ and $\deltacp$ (which, in principle, also depends on
the simulated mass hierarchy). Similar to the mass hierarchy sensitivity, we then
stack the values of $\deltacp$ to the ``Fraction of $\deltacp$''. Note, however,
that no experiment can have a CP fraction $1.0$ for the CP violation sensitivity at
any point, since there will be no CP violation sensitivity for values of $\deltacp$ close
to $0$ and $\pi$.

\subsection{Scaling with $\gamma$ and optimization}

\begin{figure}[t]
\begin{center}
\includegraphics[width=\textwidth]{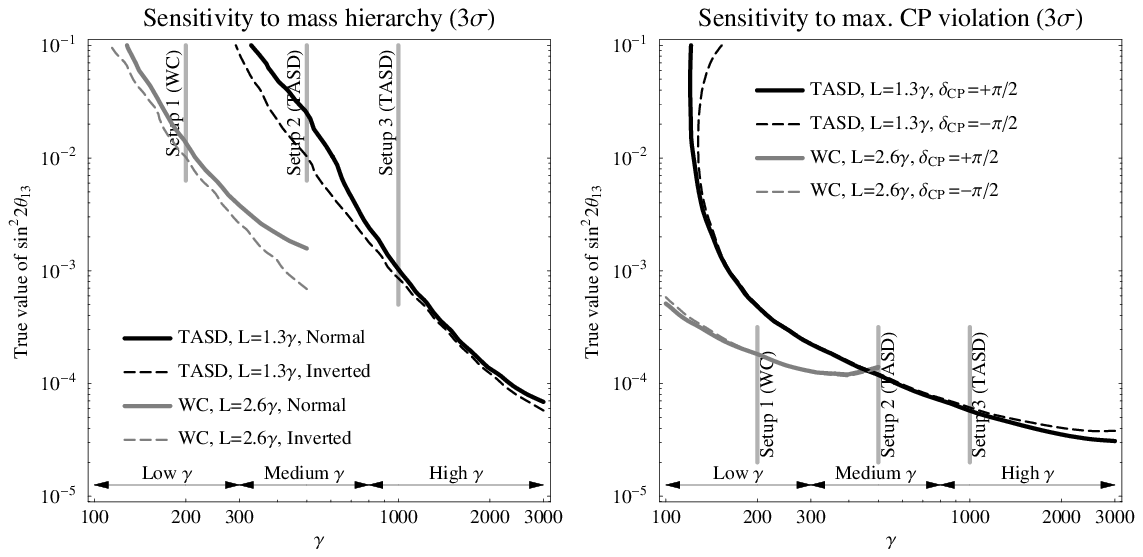}
\end{center}
\mycaption{\label{fig:mhcp} The sensitivity to the mass hierarchy (left,
  $\deltacp=0$) and maximal CP violation (right, normal hierarchy) as function 
  of $\gamma$ and the true value of $\stheta$ for the detector types and 
  parameters as described in the plot legends, where sensitivity
  at the $3\sigma$ confidence level is given above the curves.}
\end{figure}

Similar to the $\stheta$ sensitivity, one can discuss the mass hierarchy and CP violation
sensitivities as function of $\gamma$ for the different detector technologies. This
comparison is shown in \figu{mhcp} for the chosen range of $L/\gamma$. Since
higher $\gamma$ implies a longer baseline, it also implies stronger matter
effects, where we use the average density along a specific baseline. Therefore
the mass hierarchy sensitivity also improves with higher values of $\gamma$ 
(\cf, left plot). The different choice of $L/\gamma$ for the WC detector
implies that the mass hierarchy sensitivity is already present at about half
of the $\gamma$ for the TASD. There is no substantial difference between
the normal and inverted mass hierarchies, because all setups use approximately 
equal neutrino and anti-neutrino rates. For the CP violation sensitivity,
higher $\gamma$'s are, in principle, favorable, since they imply larger event 
rates. However, for very high $\gamma$, the missing energy resolution
of the non-QE events (WC) and the matter density uncertainties (TASD) act
counter-productive. For the $\beta$-beams, there seem to exist only very little
problems with degeneracies for the CP violation sensitivity, because the
measurement at the oscillation maximum helps to resolve the degeneracies.

\begin{figure}[t]
\begin{center}
\includegraphics[width=\textwidth]{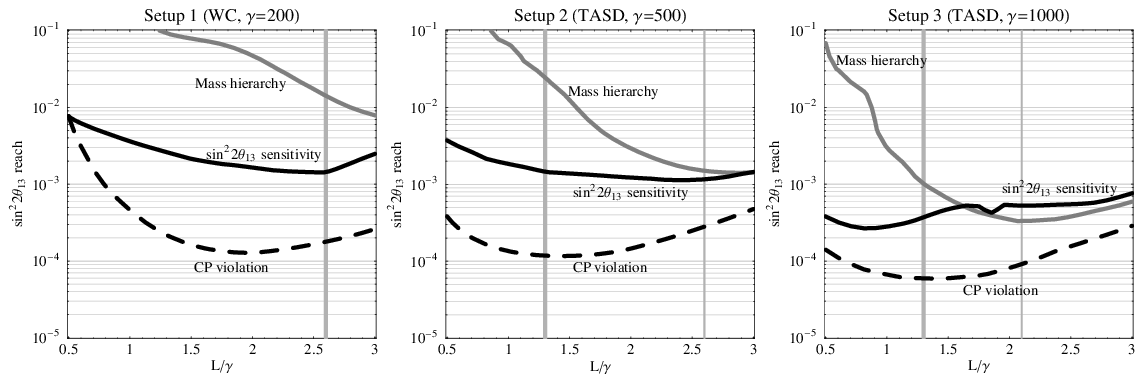}
\end{center}
\mycaption{\label{fig:lcomp} The $\stheta$ reaches for the sensitivity to
$\stheta$ (black solid curves), maximal
CP violation $\deltacp=\pi/2$ (dashed curves), and the normal mass hierarchy for
$\deltacp=0$ (gray solid curves) as function of $L/\gamma$ for the different scenarios
in the figure captions ($3 \sigma$ confidence level). The thick vertical lines correspond
to the scenarios chosen in this study. The thin vertical lines correspond to alternative
optimization strategies. All these curves include correlations and degeneracies.}
\end{figure}

It is interesting to discuss what the choices of $L/\gamma$ for Setups~1,~2, 
and~3 are really optimized for. So far, we have demonstrated that the 
choice of $L/\gamma=2.6$ for the WC detector and $L/\gamma=1.3$ for 
the TASD are quasi-optimal for the $\stheta$ sensitivity and lead to a 
clear hierarchy of these setups for the $\stheta$ discovery reach. 
We show in \figu{lcomp} the $L/\gamma$-dependence for all three setups
and the $\stheta$ sensitivity (black solid curves), maximal CP violation
sensitivity (dashed curves), and normal mass hierarchy sensitivity for 
$\deltacp=0$ (gray curves). The thick vertical lines correspond to our 
choices of $L/\gamma$, whereas the thin lines represent alternative 
optimization strategies. Setup~1 is apparently optimized for the $\stheta$ 
sensitivity, where the large $L/\gamma$ clearly favors the mass hierarchy
sensitivity and hardly affects the CP violation sensitivity. Thus, it represents
a good compromise for all quantities. Setups~2 and~3 are optimized for 
CP violation, where the $\stheta$ sensitivity is in both cases very close to 
the optimum. For Setups~2 and~3, however, the mass hierarchy sensitivity
would be considerably better close to the second oscillation maximum 
$L/\gamma \sim 2.6$, while the $\stheta$ sensitivity would be hardly
degraded. The choice of the baseline depends therefore for Setups~2 and~3
on the priorities, \ie\ if one optimizes for mass hierarchy or CP violation 
measurements. Since we also use the setup NF@7500km for the neutrino factory, 
we will show Setups~2 and~3 at the second oscillation maximum in some cases 
for a fair comparison of the mass hierarchy sensitivity.  Note that Setup~3 
could, in principle, also be operated at the ``magic baseline'', where the 
mass hierarchy and $\stheta$ sensitivities are only somewhat worse than 
optimal, but no CP violation measurement is possible. The only quantity which 
is not shown here is the $\stheta$ discovery reach. One can show that it is 
substantially degraded for the second oscillation maximum at Setups~2 or~3 
($L=2.6 \gamma$). In particular, Setup~2 would perform much worse than
Setup~1. However, the choice of the first oscillation maximum instead of 
the second would hardly change the parameter space coverage of Setup~2. 
Thus, our choices of $L/\gamma$ are consistent with the primary objective 
to discover $\stheta$.

\begin{figure}[t]
\begin{center}
\includegraphics[width=\textwidth]{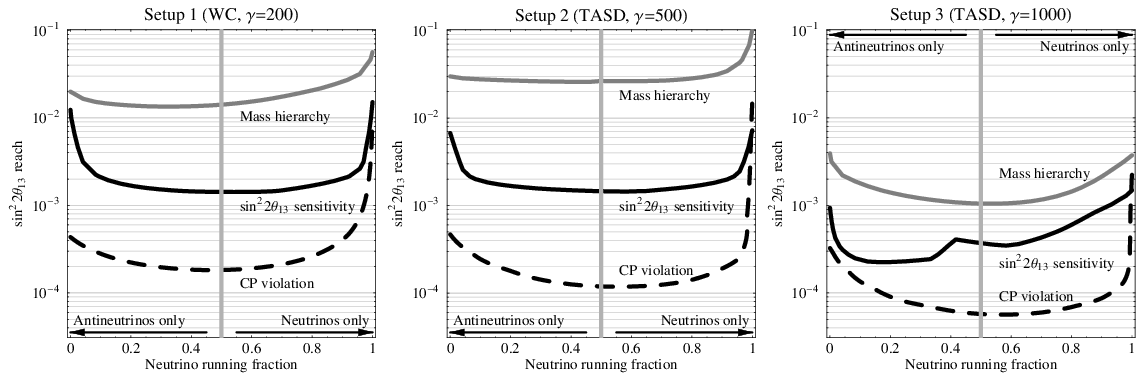}
\end{center}
\mycaption{\label{fig:rdep} The $\stheta$ reaches for the sensitivity to
$\stheta$ (black solid curves), maximal CP violation $\deltacp=\pi/2$ 
(dashed curves), and the normal mass hierarchy for $\deltacp=0$ (gray solid 
curves) as function of the neutrino running fraction for the different 
scenarios in the figure captions ($3 \sigma$ confidence level). The neutrino 
running fraction is the neutrino running time divided by the total running 
time of eight years.  The thick vertical lines correspond to the scenarios 
chosen in this study. All these curves include correlations and degeneracies 
and are computed for the normal mass hierarchy.}
\end{figure}

Recently the important issue has been raised that the neutrino event rates
might be substantially suppressed compared to the anti-neutrino event rates.
We show therefore in \figu{rdep} the dependence on the neutrino running
fraction for all three setups and the $\stheta$ sensitivity (black solid
curves), maximal CP violation sensitivity (dashed curves), and normal mass 
hierarchy sensitivity for $\deltacp=0$ (gray curves). The vertical thick lines 
correspond to our choice of 50\% neutrino and 50\% anti-neutrino running. 
The neutrino running fraction $f$ is the fraction of neutrino running divided by
the total running time of eight years, \ie\ the experiment runs $f\times8$
years with neutrinos and $(1-f) \times 8$ years with anti-neutrinos. From 
\figu{rdep}, we find that all setups are at the optimal performance for the 
CP violation measurements (vertical lines). Setups~1 and~2 also have optimal 
$\stheta$ sensitivity, whereas Setup~3 has optimal mass hierarchy sensitivity.
As far as the symmetry of the plots is concerned, running with only
anti-neutrinos is clearly favored compared to running with only neutrinos
(extreme cases), because we have somewhat higher anti-neutrino event rates 
and lower backgrounds, \ie\ the absolute rate is higher for the anti-neutrino 
case. For the inverted hierarchy (\figu{rdep} is shown for the normal
hierarchy), only running with neutrinos is even slightly more disfavored 
because of the matter suppression of the neutrino rate. It is interesting 
to note that even rather substantial deviations from a symmetric neutrino 
and anti-neutrino operation does not have extreme effects on the
measurements. Setup~3 is most affected by such deviations, where a lower
neutrino fraction means better statistics and thus a better $\stheta$ 
sensitivity, but it creates an imbalance between neutrinos and anti-neutrinos 
affecting the CP violation sensitivity. Nevertheless, it does not make sense 
to run with neutrinos or anti-neutrinos only, since this ratio would lead 
to degrading the sensitivities by an order of magnitude. Setup~1, for 
example, then looses its competitiveness compared to superbeam upgrades.
In all cases, at least 10\%-20\% of neutrino running is necessary, which 
corresponds rescaled to at least a total number of $1 \cdot 10^{18}$ useful 
$^{18}$Ne decays plus $~26 \cdot 10^{18}$ useful $^{6}$He decays.

\subsection{Comparison with other technologies}

\begin{figure}[t]
\begin{center}
\includegraphics[width=\textwidth]{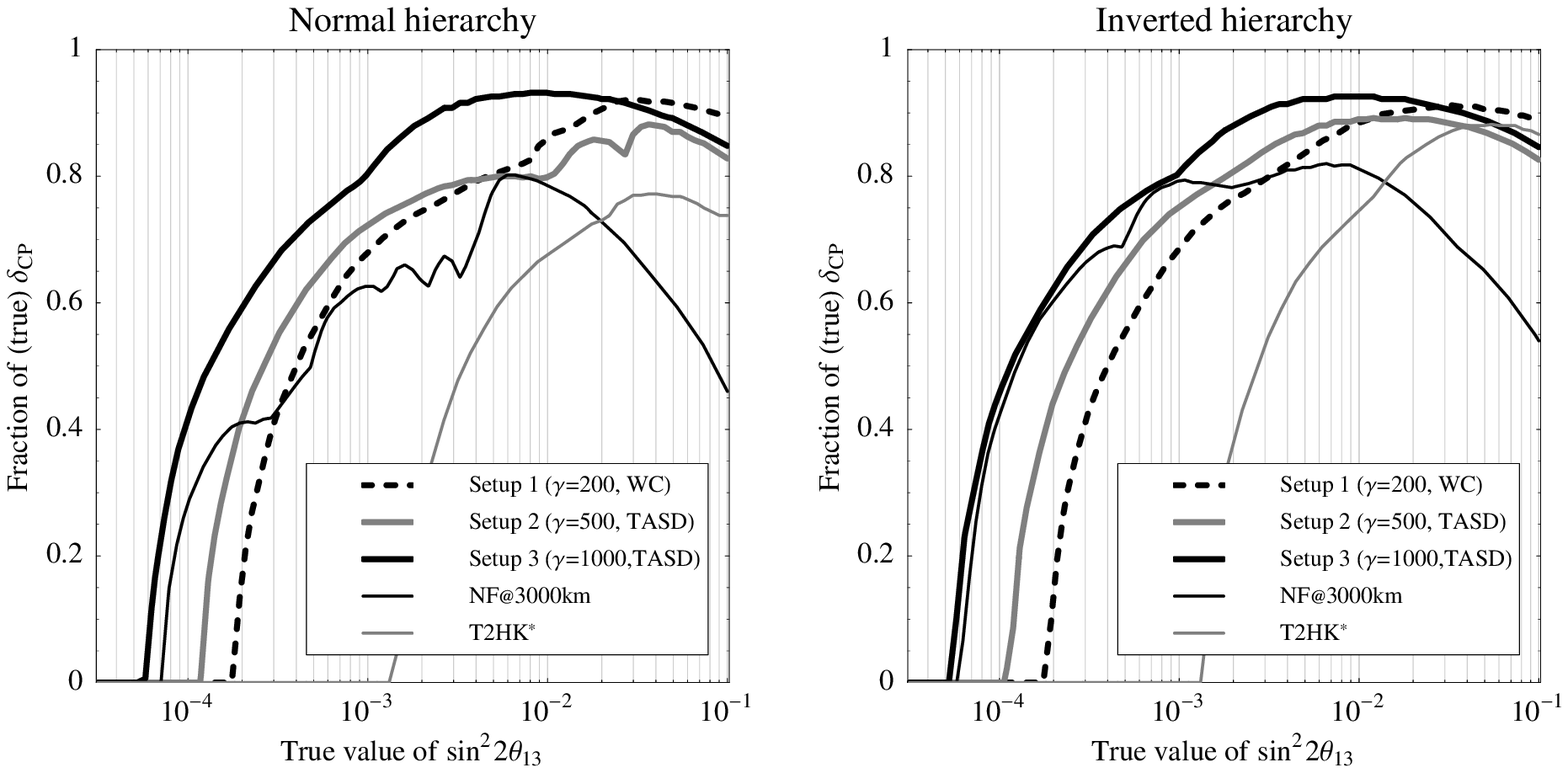}
\end{center}
\mycaption{\label{fig:cpcpf} The sensitivity to CP violation for the normal
(left) and inverted (right) mass hierarchy for different experiments as function
of the true values of $\stheta$ and $\deltacp$ at the $3 \sigma$ confidence
level. The values of $\deltacp$ are ``stacked'' to the CP fraction, which
described the fraction of all values of $\deltacp$ for which the mass hierarchy
can be resolved for a given $\stheta$. If the probability distribution is
uniform in $\deltacp$, the fraction of $\deltacp$ would directly correspond 
to the probability to measure the indicated mass hierarchy for a given $\stheta$.}
\end{figure}

In \figu{cpcpf} the sensitivity to CP violation is shown for the normal 
(left) and inverted (right) mass hierarchy for different experiments as
function of the true values of $\stheta$ and $\deltacp$ at the $3 \sigma$ 
confidence level. This figure clearly demonstrates that for a normal mass 
hierarchy all of the discussed $\beta$-beam options have an impressing 
CP violation sensitivity very competitive to the neutrino factory, 
because $(\deltacp,\stheta)$-degeneracy~\cite{Burguet-Castell:2001ez} and
``$\pi$-transit'' of the $\mathrm{sgn}(\ldm)$-degeneracy~\cite{Huber:2002mx} 
destroy the CP violation sensitivity of the neutrino factory at many places. 
Note that these degeneracy problems could, in principle, be resolved by a 
combination with the magic baseline, but a much better sensitivity than 
that of Setup~3 is unlikely to be achieved. As far as the $\stheta$
reach is concerned (in horizontal direction), there is again a clear 
hierarchy among Setups~1,~2, and~3. For large values of $\stheta$, however, 
matter density uncertainties affect the longer baselines, and Setup~2 has 
to deal with some problems due to degeneracies (left plot). For optimal 
$\stheta$, Setup~3 can establish CP violation for more than 90\% of all 
values of $\deltacp$, whereas  the neutrino factory is limited to about 
80\%. For the inverted hierarchy, the matter effects enhance
the anti-neutrino rate, which means that the neutrino and (lower) 
anti-neutrino rates at the neutrino factory are getting more equal statistical 
weight and the correlations can easier be resolved. Balanced event rates
of the $\beta$-beams lead, on the other hand, to very little impact of 
the mass hierarchy. For \JHFHK , the somewhat lower anti-neutrino
rate implies a similar behavior to the neutrino factory.

\begin{figure}[t]
\begin{center}
\includegraphics[width=\textwidth]{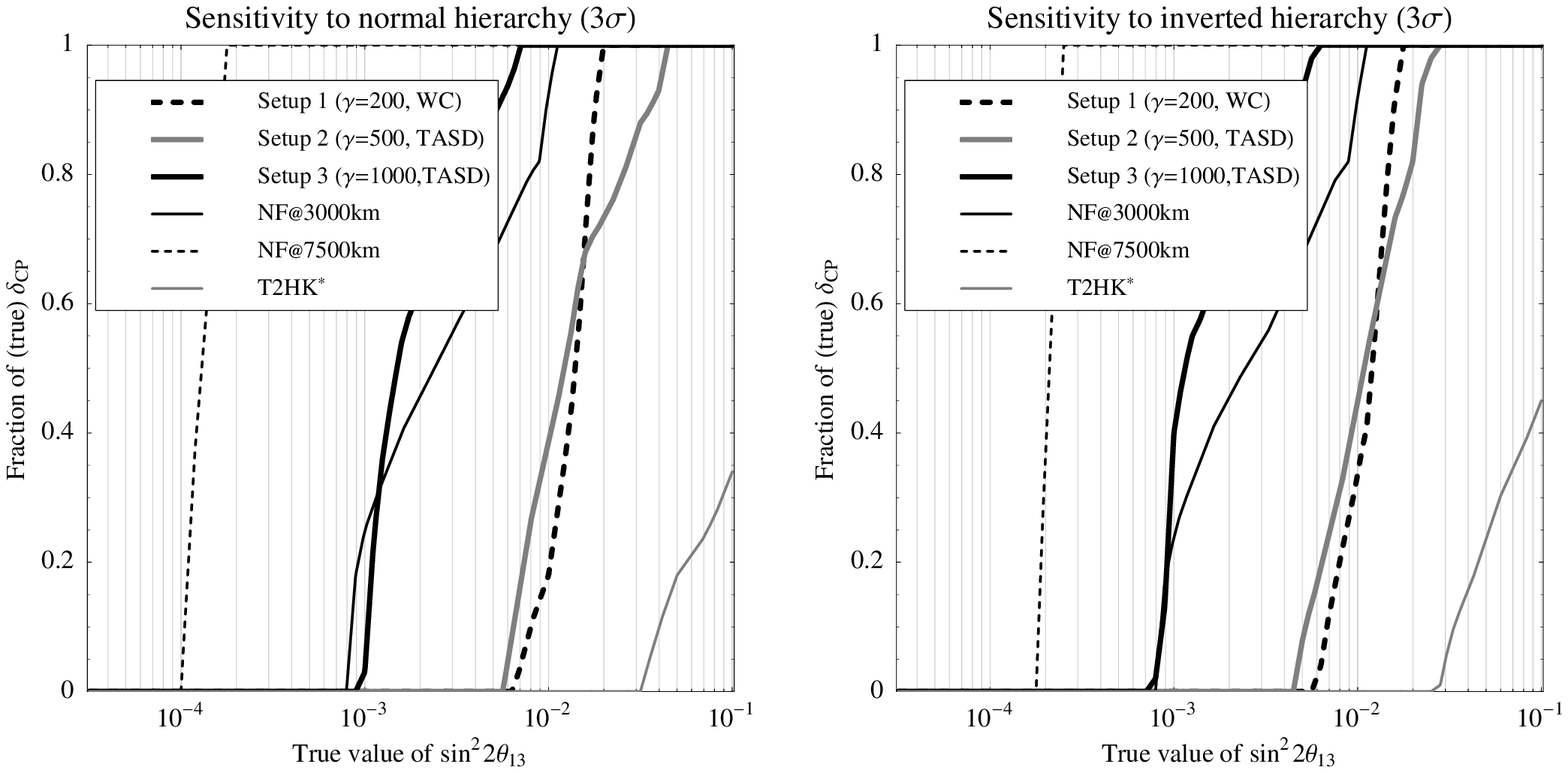}
\end{center}
\mycaption{\label{fig:signcpf} The sensitivity to a normal (left) and 
inverted (right) mass hierarchy for different experiments as function 
of the true values of $\stheta$ and $\deltacp$ at the $3 \sigma$ confidence 
level. The values of $\deltacp$ are ``stacked'' to the CP fraction, 
which describes the fraction of all values of $\deltacp$ for which the 
mass hierarchy can be resolved for a given $\stheta$. If the probability 
distribution was uniform in $\deltacp$, the fraction of $\deltacp$ 
would directly correspond to the probability to measure the indicated 
mass hierarchy for a given $\stheta$.}
\end{figure}

In order to compare the mass hierarchy sensitivity of all options, we show 
in \figu{signcpf} the mass hierarchy sensitivity as function of $\stheta$ 
and $\deltacp$ for the normal (left) and inverted (right) mass hierarchy 
for the Setups defined in the last section and the neutrino factory and 
superbeam representatives. Given the choice of $L/\gamma$, Setups~1
and~2 have a very similar mass hierarchy sensitivity because of the very 
similar baselines, whereas Setup~3 is substantially better. In all cases, 
the neutrino factory at the magic baseline covers by far the most parameter 
space, whereas the performance of NF@3000km is very close to the one of 
Setup~3. The superbeam setup \JHFHK\ can only establish the mass hierarchy
for a very small fraction of $\deltacp$ because of its short baseline. 
Note, however, that other superbeam upgrades (such as FNAL-Homestake 
or BNL-Homestake) could have a much better mass hierarchy
sensitivity~\cite{Diwan:2002xc}. 
The relative performance of the neutrino factory baselines is degraded 
for the inverted versus normal hierarchy, because the event rates are 
not evenly distributed between neutrinos and anti-neutrinos. As we have 
discussed in the last section, a longer baseline for Setups~2 and~3 
improves the mass hierarchy sensitivity drastically. It turns out that 
for $L/\gamma=2.6$, Setup~2 is comparable to the neutrino factory at 3000~km, 
whereas Setup~3 is almost as good as the neutrino factory at the magic baseline.

\begin{figure}[t]
\begin{center}
\includegraphics[width=0.70\textwidth]{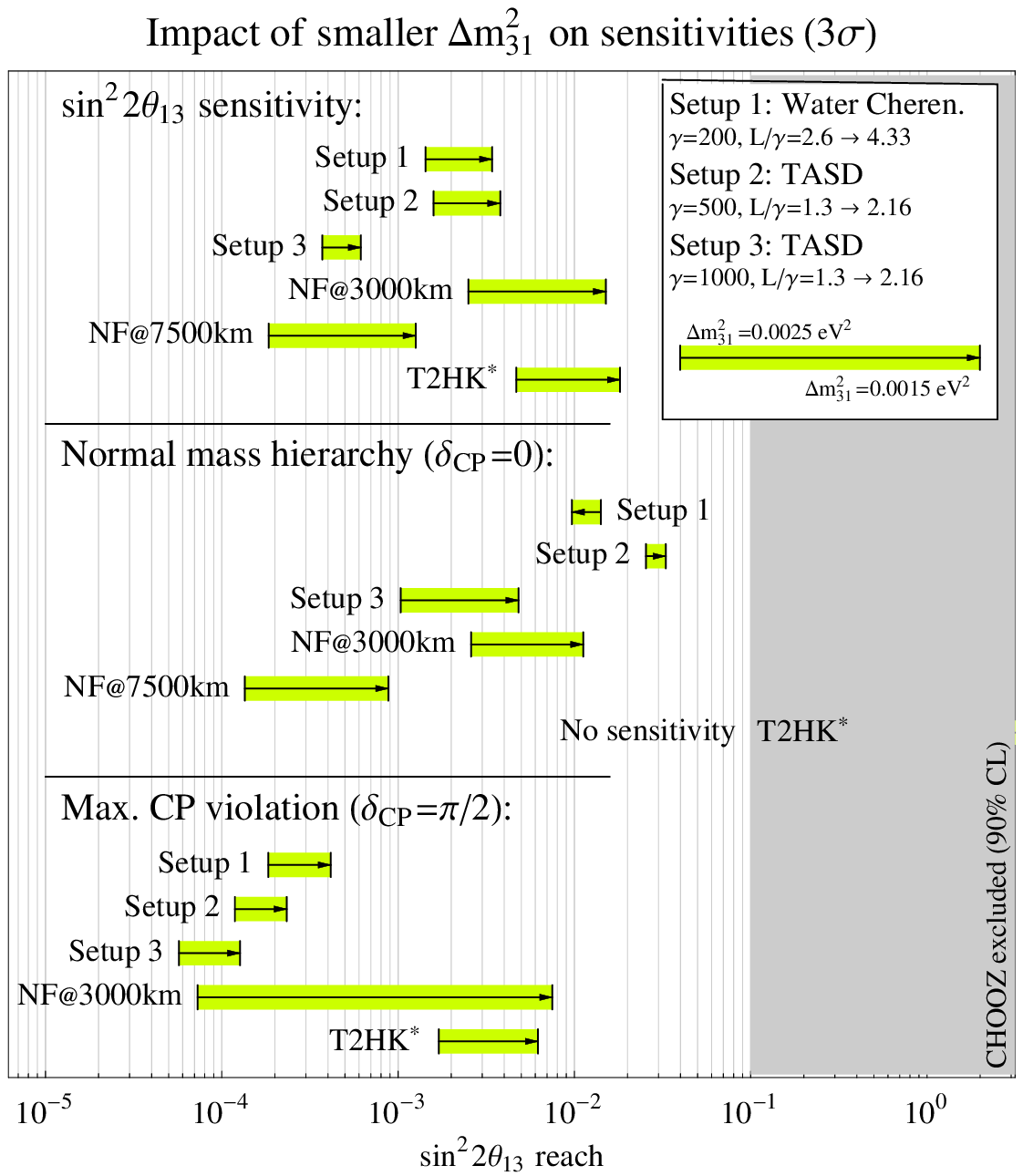}
\end{center}
\mycaption{\label{fig:lowldm} Impact of a smaller value of $\ldm$ on different 
sensitivities of various experiments including all correlations and degeneracies
($3\sigma$ confidence level). The bars represent the shift from 
$\ldm = 0.0025 \, \mathrm{eV}^2 \rightarrow \ldm = 0.0015 \, \mathrm{eV}^2$
(simulated values), where the $\beta$-beam baselines are re-scaled according to 
$L \rightarrow L \times 0.0025/0.0015$. Note that for the CP violation
sensitivity in this figure, we have chosen the smallest value of $\stheta$ for 
which maximal CP violation could be established.}
\end{figure}

A very interesting (though not very likely) case for the $\beta$-beams could 
occur if $\ldm$ turns out to be at the lower end of the currently allowed 90\% CL 
region~\cite{Maltoni:2004ei},
\ie, $\ldm \simeq 0.0015 \, \mathrm{eV}^2$. Since $\ldm$ will be measured 
to a high precision soon by MINOS, T2K, and NO$\nu$A, it is straightforward 
to re-optimize the $\beta$-beams by just moving the detector back into the 
oscillation maximum. For the neutrino factory, however, the oscillation 
maximum for the mean energy is at a very long baselines 
$L \gg 7\, 500 \, \mathrm{km}$ anyway, which means that moving the 
detector to an even longer baseline should have an effect similar to 
choosing the magic baseline scenario directly. In addition, other constraints 
may prevent the selection of longer baselines.
We illustrate the effect of a smaller $\ldm$ in \figu{lowldm}, where the arrows
indicate the shift. In this figure, the $\beta$-beam baselines are re-scaled
according to $L \rightarrow L \times 0.0025/0.0015$ in order to stay in the oscillation
maximum, whereas the other baselines are fixed. In almost all cases the experiments
loose sensitivity. However, the relative shift for the neutrino factories is in some
cases much larger because the smaller $\ldm$ means that the oscillation peak is
shifted to lower energies where the charge identification requirement leads to lower
efficiencies. In particular, the CP violation sensitivity of the neutrino factory
is highly affected. For the mass hierarchy sensitivity, the neutrino factory at the
magic baseline is still the best experiment. Note that Setups~1 and~2 are hardly affected by
the different value of $\ldm$, since a smaller value of $\ldm$ means a longer baseline
and the stronger matter effects partially (Setup~2) or fully (Setup~1)
compensate for the drop in $1/L^2$. To be fair, this comparison is only shown for selected
parameter choices and for the assumption of unflexible neutrino factory baselines.
Indeed, for a smaller value of $\ldm$, a dedicated study is required which re-optimizes
all potential experiments. However, one can easily see from this figure that the
charge identification cuts at low energies for the neutrino factory imply that one
quickly ends up at inconveniently long baselines for such an optimization. This
behavior is not expected for the $\beta$-beams.

\section{Summary and conclusions}
\label{sec:summary}

In this study, we have discussed various optimization aspects of
$\beta$-beams and we compared the physics potential to neutrino
factories and superbeam upgrades. Two central parameters are the
gamma factor and the baseline of the $\beta$-beam which were taken
to be free parameters. We considered two different detector
technologies in connection with the $\beta$-beam, namely a Water
Cherenkov detector and a Totally  Active Scintillator Detector
with $500 \, \mathrm{kt}$ and  $50 \, \mathrm{kt}$ fiducial mass,
respectively. The Water Cherenkov detector was also considered as
a target for a superbeam upgrade which we called \JHFHK . For the
comparison with the neutrino factory we used a $50 \, \mathrm{kt}$
magnetized iron detector at baselines of 3000~km and 7500~km.
An important aspect concerns the assumptions about the number
of ion decays per year. One scenario was to assume that the
number of decays does not depend on $\gamma$. However, this is
for a number of reasons technologically not realistic and we
studied therefore scenarios where the number of decays per year
scales with $\gamma$ like a power law. For the superbeam upgrade
and  the neutrino factory ``standard'' beam luminosities were
assumed  (see section \ref{sec:comparetech} for details).
As performance indicators, we have considered the $\stheta$
sensitivity (corresponding to the exclusion limit which can be
achieved by an experiment), the $\stheta$ discovery reach, the
(normal and inverted) mass hierarchy discovery reach, and the
CP violation discovery reach for both hierarchies. Specific
experimental setups were defined to allow a more detailed comparison.

\begin{figure}[t]
\begin{center}
\includegraphics[width=0.72\textwidth]{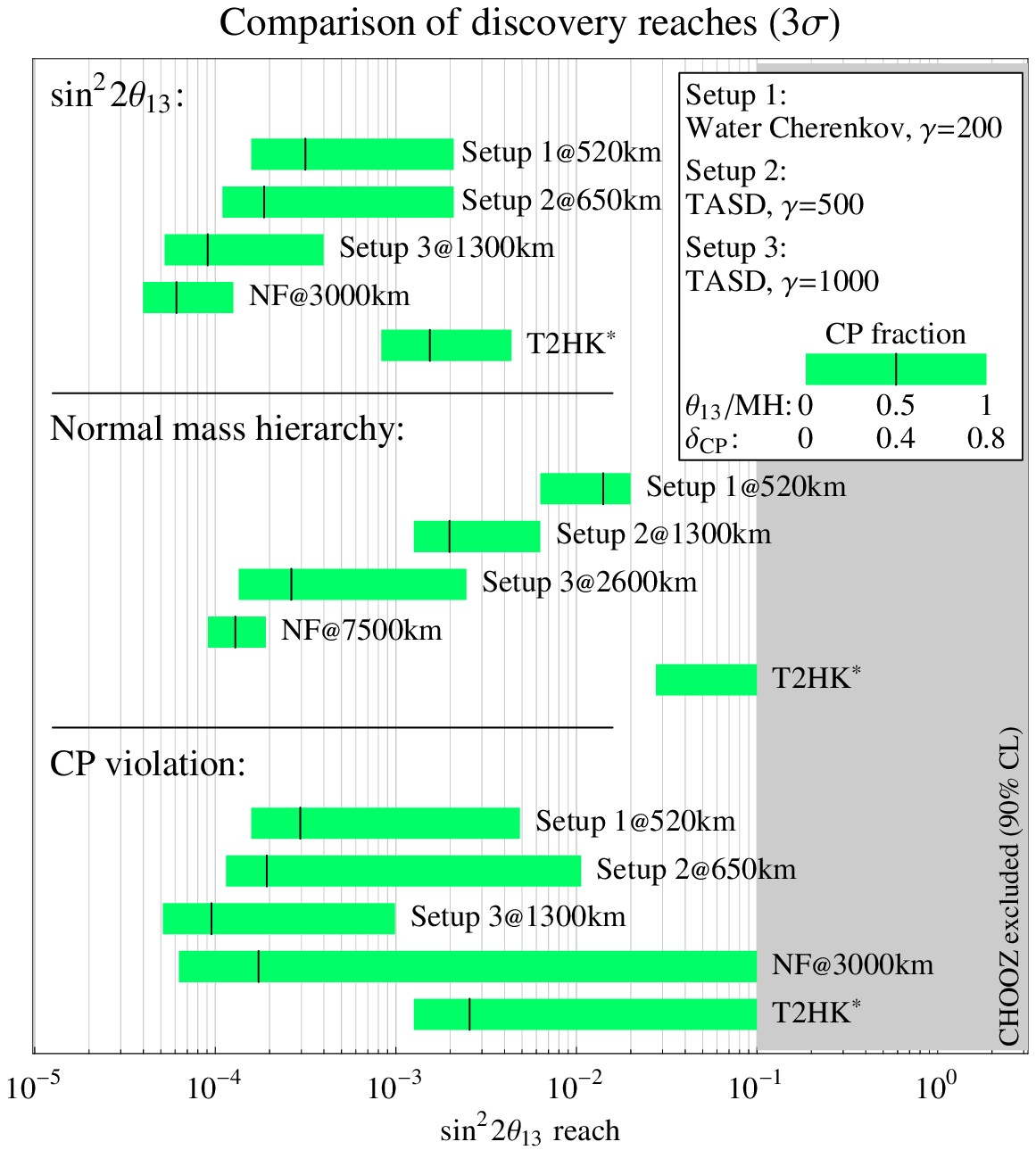}
\end{center}
\mycaption{\label{fig:discsummary} Discovery reach comparison at the $3\sigma$ 
confidence level. The bars represent the best case (left end), ``typical
case''  (middle line), and worst case (right end) in $\deltacp$, where the 
respective sensitivities are computed with the CP fractions in the plot
legend.  Note that for CP violation, a CP fraction of one can never be 
achieved since values close to $0$ or $\pi$ cannot be distinguished from CP 
conservation. Within each category, the most competitive setups from this 
study are compared (first or second oscillation maximum for $\beta$-beams). 
For all shown sensitivities, a normal mass hierarchy is assumed.}
\end{figure}

Our main results for the discovery reaches are summarized in
\figu{discsummary}, where the bands reflect the impact of the
true value of $\deltacp$.
We find that the choice of the optimal $\gamma$ clearly depends on the
objectives of the $\beta$-beam experiment and external constraints.
We in general find good agreement with existing studies showing that
$\gamma$ should be at least high enough to avoid the Fermi-motion
dominated regime in order to have sufficient energy 
information~\cite{Burguet-Castell:2003vv,Burguet-Castell:2005pa}. 
Low $\gamma \lesssim 300$ can be achieved with relatively ``small''
accelerators, such as of SPS size in combination with Water Cherenkov
detectors. All of the discussed performance indicators imply that
a $\beta$-beam in this range clearly outperforms the \JHFHK\ superbeam
upgrade using the same detector since it is not limited by the intrinsic
beam background. In fact, the CP violation discovery reach is already
quite close to the optimum even compared with higher gamma or neutrino
factory setups. Note that this excellent simultaneous sensitivity to
$\stheta$, mass hierarchy, and CP violation is achieved by including
the second oscillation maximum to disentangle correlations and
degeneracies. An operation at a shorter baseline would significantly
affect the final mass hierarchy and $\stheta$ sensitivities, and it would
hardly help the CP violation sensitivity. Note that we have chosen
the ion decay rates such that there are approximately equal neutrino
and anti-neutrino event rates for all setups. This balanced concept
implies that the $\beta$-beam performance hardly depends on the mass
hierarchy and the performance for CP violation is excellent. However,
if the neutrino rate were significantly lower than the anti-neutrino
rate, then the CP violation sensitivity would be affected.

The range $300 \lesssim \gamma \lesssim 800$, requires already relatively 
large accelerator rings, such as of Tevatron size. As detector technology, 
we have chosen the Totally Active Scintillator Detector (TASD) in this case, 
because it is rather predictable in this gamma-range. However, it is not
excluded that the extrapolation to a ten times larger Water Cherenkov 
could result in a better performance. For the setup in this and the 
larger gamma range, we find that the choice of $L/\gamma=1.3$ or $2.6$ 
clearly depends on the optimization goals. For $\stheta$ discovery and 
CP violation potential, shorter baselines are favored, whereas sensitivity
to the mass hierarchy favours longer baselines. The $\stheta$ is rather 
indifferent with respect to the baseline choice, since statistics 
becomes worse for longer baselines, while the matter effects become
stronger and the second oscillation maximum helps to resolve correlations 
and degeneracies. Except for CP violation, even the optimized medium 
$\gamma$ setup is not competitive to the neutrino factory, but it could 
be an interesting step towards it.

Large $\gamma \gtrsim 800$ would require very large and powerful 
accelerators of LHC size, where we use TASD as detector technology.
We find $\stheta$ (sensitivity and discovery) and mass hierarchy discovery 
reaches close to the neutrino factory setups, and a CP violation discovery 
reach better than the one of the neutrino factory setups
(\cf, \figu{barscomp} and \figu{discsummary}). In this case, a $\beta$-beam 
could clearly be a competitor to a neutrino factory, if technically feasible 
for such a high $\gamma$. 

We have assumed initially that the number of ion decays per year is constant 
in $\gamma$ and we investigated the impact of external constraints to this 
assumption. We found that especially the high gamma setup suffers from 
modifications in the scaling of ion decays. This implies that the ion 
luminosity is a critical factor for this setup. In addition, we compare in 
\figu{discsummary} setups optimal for the individual purposes, \ie,
differently optimized $\beta$-beams with correspondingly optimized neutrino factories.
If we required optimal sensitivity to all quantities, we would need two 
neutrino factory or two $\beta$-beam baselines. Though a storage ring with 
two decay sections could, in principle, be possible for a $\beta$-beam, 
it might be more compact in the neutrino factory case. If, however, one 
of the measurements becomes obsolete, such as the mass hierarchy which 
might be determined by a supernova explosion, the $\beta$-beam baseline 
optimization seems to be quite straightforward, whereas the neutrino factory 
may still require another baseline to resolve degeneracies. For a precision 
measurement of $\deltacp$, for example, the ``magic baseline'' could be
required for unfortunate values of $\deltacp$~\cite{Huber:2004gg}.

The $\beta$-beam has in all cases the advantage that the baseline can be 
freely chosen such that one is measuring at the oscillation maximum. 
For the case of a much smaller $\ldm$ than the current best-fit value, 
first tests suggest that the performance improves compared 
to neutrino factories. However, this less likely case requires further
study because one should also optimize the neutrino factory for such a 
different value of $\ldm$.

We conclude that a lower gamma $\beta$-beam is certainly an interesting 
physics alternative to a large superbeam upgrade and a higher gamma 
$\beta$-beam could be an competitive alternative to a neutrino factory. 
In all cases, the attractiveness of the $\beta$-beams depends clearly 
on the ability to produce enough isotope decays for both neutrinos and 
anti-neutrinos. Especially for $\deltacp$ measurements, the $\beta$-beams 
might then outperform all existing techniques, whereas for $\stheta$ 
discovery and mass hierarchy sensitivity the neutrino factory is ultimately 
a better choice. Except from $\stheta$, mass hierarchy, and CP violation
measurements, there is more physics to be done with a neutrino factory. 
In the neutrino oscillation sector, for example, there is sensitivity to 
the leading atmospheric parameters. The case $\stheta=0$ would suggest 
to use the  $\nu_\mu$ disappearance channel at a very long baseline for 
mass hierarchy and MSW effect measurements~\cite{Winter:2004mt,deGouvea:2005hk}. 
This role is unlikely to be replaced by the electron neutrino disappearance 
channel of a $\beta$-beam. Nevertheless, we conclude that $\beta$ beams 
constitute a very interesting option for future precision neutrino 
oscillation experiments. Further technological feasibility studies are 
clearly well motivated to explore if a $\beta$ beam can be realized. The 
technical feasibility, the financial effort, and the physics potential of 
a $\beta$-beam and a neutrino factory have to be compared then again before 
an ultimate decision is made.

\subsubsection*{Acknowledgments}

We would like to thank M.~Lindroos, P.~Litchfield, M.~Mezzetto, and L.~Mualem 
for useful information. This work has been supported by SFB 375 and the 
Graduiertenkolleg 1054 of Deutsche Forschungsgemeinschaft. WW would like to 
acknowledge support from the W.~M.~Keck Foundation and NSF grant PHY-0070928. 
In addition, he would like to thank the theory groups at TUM and Wisconsin for 
their warm hospitality during his visits, where parts of this work
have been carried out.

\begin{appendix}

\newpage

\section*{Appendix A: Water Cherenkov detector and 2nd oscillation maximum}
\label{app:2ndosc}

\begin{figure}[tbp]
\begin{center}
\includegraphics[width=0.7\textwidth]{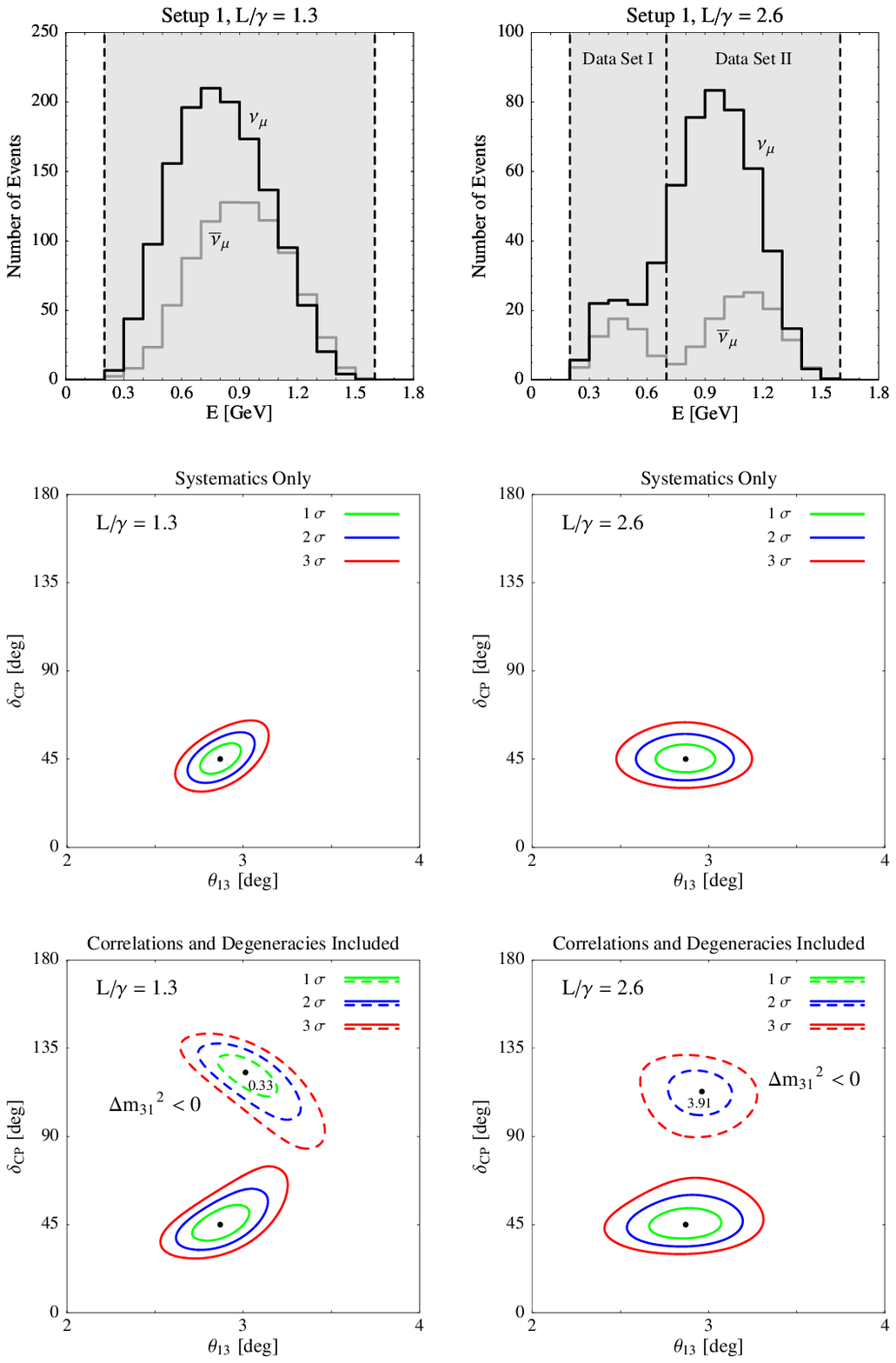}
\end{center}
\mycaption{\label{fig:1vs2_Panel} Comparison of the performance of Setup~1 at
L/$\gamma$=1.3 (left column) and L/$\gamma$=2.6 (right column). The oscillation
parameters used as input are $\sin^22\theta_{13}=10^{-2}$, $\delta=\pi/4$ and
the other parameters are the ones from \equ{standard_params}.
The first row shows the appearance spectra for neutrinos and anti-neutrinos in
the energy window of the analysis (gray area). The second row shows the allowed
regions in the $\theta_{13}$-$\delta$ plane at 1, 2, and 3 $\sigma$ with only
taking systematics into account, while in the third row also correlations and
degeneracies are included. The value next to the local minimum in the degenerate
solution (fitted with $\Delta m_{31}^2<0)$ indicates the value of $\Delta
\chi^2$ at the local minimum.}
\end{figure}
The better performance of Setup~1 at the longer baseline with $L/\gamma = 2.6$
(although statistics drops with $1/L^2$) can be understood with \figu{1vs2_Panel}
where the comparison of $L/\gamma = 1.3$ (left column) and $L/\gamma = 2.6$ (right
column) is shown. The appearance spectra for $\sin^22\theta_{13}=0.01$ and
$\delta_{CP}=\pi/4$ is shown in the first row of \figu{1vs2_Panel} for neutrinos
(black curve) and anti-neutrinos (gray curve).
For Setup 1 at $L/\gamma = 1.3$ one can clearly see, that only the first
oscillation maximum contributes to the whole appearance spectra, while for the one
at $L/\gamma = 2.6$ the first oscillation appearance maximum is shifted to higher
neutrino energies and appearance events from the second oscillation maximum enter
the energy window of the analysis from lower energies but the overall event rates
are decreased. In the second row,
we show the allowed regions in the $\theta_{13}$-$\delta$ plane at 1, 2, and 3
$\sigma$ for the same true oscillation parameters (indicated by the black dot)
where only systematical errors are taken into account. The allowed regions for the
$L/\gamma = 2.6$ scenario are somewhat larger due to the lower statistics. But only
if also correlations and degeneracies are included (third row) one can see the
impact of the second oscillation maximum. The degenerate solution fitted with
$\Delta m_{31}^2<0$ is smaller for the $L/\gamma = 2.6$ scenario and does not
reach to higher values of $\theta_{13}$ than the region that contains the best-fit
value. Additionally the $\Delta \chi^2$ value at the local minimum of the
degenerate solution is much higher with $L/\gamma = 2.6$ than the one with
$L/\gamma = 1.3$ and does not even appear at 1 $\sigma$.

\begin{figure}[t]
\begin{center}
\includegraphics[width=0.9\textwidth]{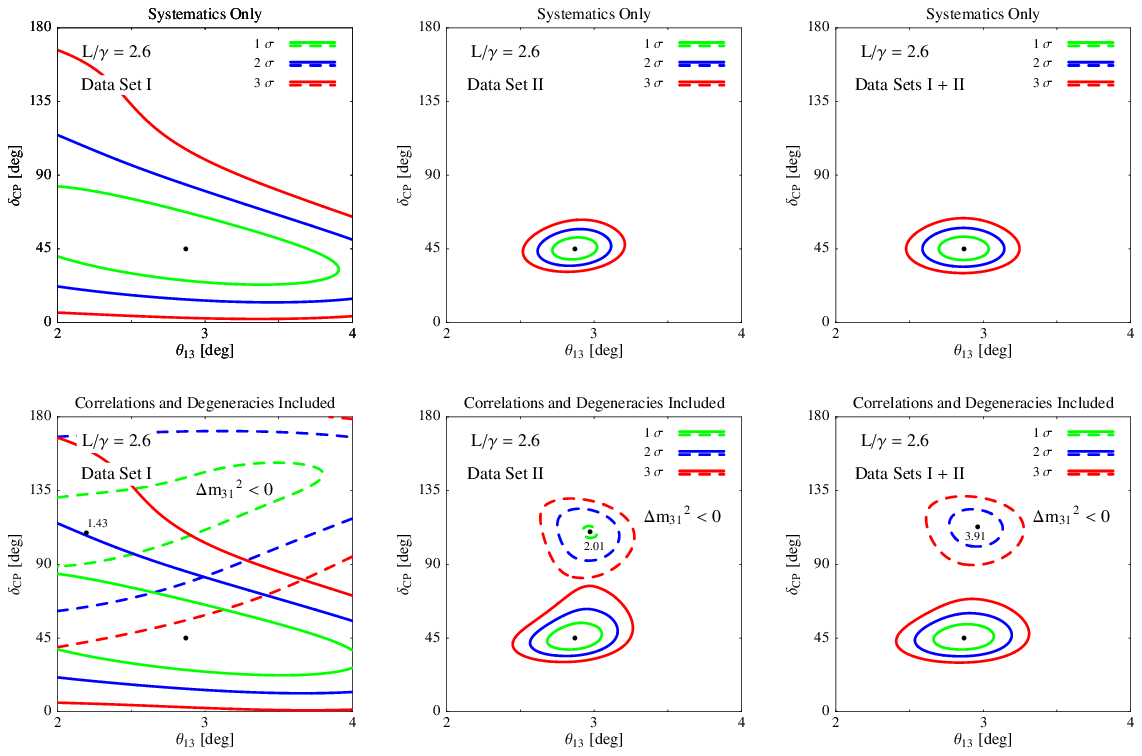}
\end{center}
\mycaption{\label{fig:Data1vsData2} The allowed regions in the
$\theta_{13}$-$\delta$ plane at~1,~2, and~3 $\sigma$ for Setup-I with L/$\gamma$=2.6.
In the upper row only systematics are taken into account while in the lower row
also correlations and degeneracies are included. The allowed regions are shown
for the separated data sets~I (first column) and~II (second column) as explained in the text
(see also Fig.~\ref{fig:1vs2_Panel}). In the third column the allowed regions for
the combination of both data sets is shown. The value next to the local minimum
in the degenerate solution (fitted with $\Delta m_{31}^2<0$) indicates the
value of $\Delta \chi^2$ at the local minimum.}
\end{figure}

In order to understand better the impact of the appearance events from the second
oscillation maximum in the $L/\gamma = 2.6$ scenario, we divided the whole data
set of Setup 1 in two separate data sets which only contain the appearance
events
from the first or second oscillation maximum. Data set I reaches from 0.2 to 0.7~GeV
and data set
II reaches from 0.7 to 1.6 GeV as can be seen in the upper right picture of
\figu{1vs2_Panel}. In \figu{Data1vsData2} we compare again the allowed regions
in the $\theta_{13}$-$\delta$ plane at 1, 2, and 3 $\sigma$, now for data set I
(left column), data set II (middle column) and both data sets combined (right
column). 
In the first row we only consider systematics and in the second row
correlations and degeneracies are switched on. One can clearly see that due to extremely
low statistics with only the appearance events from data set 1 the allowed regions
are strongly expanded. The allowed regions from data set II are highly improved
and in the case "systematics only" even somewhat smaller than for the combination of both
data sets. This effect comes from the fact that most of the background events
reconstruct at smaller energies (i.e. within data set I) and therefor the S/B ratio
is smaller for data set I only.
But for the degenerate
solution fitted with $\Delta m_{31}^2<0$ the combination of both data sets results
in an improvement since the local degenerate minimum for data set II lies at a
different point in the parameter plane than the one for data set I.

\end{appendix}


\clearpage
\newpage

\end{document}